\documentclass[final , letterpaper,12 pt] {report}%{scrartcl}
\usepackage[english]{babel}
\usepackage[usenames,dvipsnames]{color}
\usepackage{color}
\usepackage{graphicx}
\usepackage{amsmath}
\usepackage{indentfirst}

\usepackage{type1cm} 
\usepackage{lettrine}
\usepackage{fancyhdr}
\usepackage[titles]{tocloft}
\usepackage{multirow}
\usepackage{rotating} 
\usepackage{dcolumn}
\usepackage{setspace}
\usepackage[margin=1in]{geometry}

\title{\bf{Increasing the Discovery Power and Confidence Levels \\of Disease Association Studies:\\A Survey\\  
 \Large{\textit{\\Depth Report}}}}
\author{\\ \\ \\ \\ \textbf{\LARGE{Layan Nahlawi}}\\ \\ \\ \\ \\  \textrm{\textbf{\textit{Supervisors:}}}\\ Dr. Parvin Mousavi, Dr. Hagit Shatkay\\ \textrm{\textbf{\textit{Supervisory Committee:}}}\\Dr. Dorothea Blostein, Dr. James Stewart\\ \\ \\ \\}

\date{School of Computing, Queen\rq{}s University\\ Kingston, Ontario, Canada \\ \today}

\begin{document}
\maketitle
\thispagestyle{empty}
\pagebreak
 \renewcommand{\contentsname}{Table of Contents}
\renewcommand{\chaptermark}[1]{\markboth{\thechapter.\space#1}{}} 
\fancypagestyle{plain}{%
  \fancyhf{}\fancyfoot[R]{\thepage}%
  \renewcommand{\headrulewidth}{0pt}}
\renewcommand{\sectionmark}[1]{\markright{#1}}
\pagebreak

\pagestyle{fancy}
\fancyhead{} % clear all header fields
\fancyhead[L]{\scriptsize{\nouppercase{\rightmark}}}
\fancyhead[R]{\scriptsize{\nouppercase{\leftmark}}}
\fancyfoot{} % clear all footer fields
\fancyfoot[LE,RO]{\thepage}
\pagenumbering{roman}
\tableofcontents
\pagebreak

\listoffigures
\pagebreak

  \listoftables 
\pagebreak

\clearpage
\pagenumbering{arabic}

\chapter{Introduction}

The majority of common terminal diseases are influenced by multiple genetic and environmental factors \cite{Motulsky2006, Comings2001, Schork1997}. Cancer, Multiple Sclerosis, Alzheimers disease and Diabetes are a few examples of such diseases \cite{Gibbs2003, Bertram2007, McElroy2008, Ponder2001,Thomas2009}. These diseases are known as polygenic due to the additive and interactive contribution of multiple genes in the disease risk as opposed to monogenic diseases affected by a single gene \cite{Comings2001, Wilkie2001}.  Even though uncovering the main causes of disease is deemed difficult due to the complexity of gene-gene and gene-environment interactions, major research efforts aim at identifying disease risk factors, especially genetic ones \cite{Hemminki2006, Pearson2008, Emslie2001, Wallace2009}. 

Over the past decade, disease association studies have been used to uncover the susceptibility, aetiology and mechanisms of action pertaining to common diseases \cite{Ku2010, McCarthy2008, McCarthy2008b}. In disease association studies, genetic data is analyzed in order to reveal the relationship between different types of mutations on the DNA strands, known as variants, and a disease of interest  \cite{Wallace2009, McCarthy2008}. The ultimate goal of association studies is to facilitate susceptibility testing for disease prediction, early diagnosis and enhanced prognosis \cite{Motulsky2006, Pearson2008}. Susceptibility testing  and disease prediction  are particularly important for diseases that can be prevented by diet, drugs or change in lifestyle \cite{Motulsky2006}. The discovered associations assist in understanding  the molecular mechanisms influenced by the reported variants, and in identifying important risk factors \cite{McCarthy2008b}. Moreover, genetic risk factors are used to reveal the correlations of common variants with functional genes and their coded proteins. These correlations are in turn utilized to uncover possible target loci for better therapeutic treatments \cite{Pearson2008}. 

Different types of disease association studies analyze different types of data, genetic and non-genetic. However, the most common form of association studies utilizes genetic common variant data, specifically single nucleotide polymorphism (SNP) data \cite{Lewis2012}. A SNP is a mutation in a single nucleotide that appears in at least 1\% of the population \cite{Gibbs2003, Brookes2001}. The hypothesis behind using SNP data in association studies is that common variants influence common disorders \cite{Hemminki2008}. Additionally, common variants are easier to sequence and  analyze due the higher possibility of being encountered in the assayed population. As a consequence, common variants are involved in the majority of reported disease associations.

Current association studies suffer from several shortcomings. First, the reported associations of common variants do not explain all the variability in the genetics of complex disease. Therefore, common variant's contribution in the variability of a disease, known as its effect size on disease risk, is considered to be modest \cite{McCarthy2008b, Park2011}. The unexplained portion of the variability in disease genetics, known as the \lq\lq{}missing heritability\rq\rq{}, is believed to exist in associations of rare variants that are hard to capture with the current association studies data \cite{McCarthy2008b, Ji2008}. Second, a pressing issue in association studies is the reproducibility and replication of results in both the studied population (intra-population)  and other populations (inter-population) \cite{McCarthy2008, McCarthy2008b}.  Populations of European ancestry have been extensively analyzed in disease association studies and a significant number of informative associations was reported \cite{McCarthy2008}.  In order to generalize such findings and use them in further analysis, reproducing and replicating the results on both intra-population and inter-population basis is needed \cite{McCarthy2008}. Third, statistical hypothesis testing is used to make inferences about the association of variants with the disease being studied. The statistical significance of these inferred associations is reported using \textit{p-values}, which reflect the rate of false positive associations \cite{McCarthy2008b, Lewis2012}.  Using such methodology of testing and significance evaluation does not convey the statistical power of the study as well as the false discovery rate (the number of reported associations that are actually unassociated) \cite{McCarthy2008b, Storey2003, Nguyen2009, Berger1987}. The statistical power of a study is its ability to detect associations that are actually real and it is affected by several factors such as the sample size, the sampling distribution mean and standard deviation.

Recently, considerable research efforts have been directed toward providing strategies to overcome the challenges of association studies mentioned above. Some of these efforts focus on machine learning techniques to detect disease associations such as using decision trees and random forests \cite{Moore2010}. 
Others aim to enhance existing testing and evaluation methods for disease association studies, for instance by substituting the use of \textit{p-values} by  q-values to account for false discovery rate \cite{Storey2003}. Yet other research efforts propose techniques for post-processing the findings of association studies or complementing these findings with additional information \cite{Cusanovich2012, Hsu2010, Baranzini2009, Bush2009}. Such techniques suggest the integration of functional information, gene expression or pathway data with association studies \cite{Cusanovich2012, Hsu2010, Baranzini2009, Bush2009}. These latter techniques attempt to answer questions about disease mechanisms \cite{Ioannidis2009}. Finally, research about alternative methods for association studies seeks to find non-genetic factors in disease susceptibility such as the processes that modify the function of genes without altering the nucleotide sequence itself \cite{Rakyan2011, Lawrence2005}.
\begin{table}[!t]
\centering
\begin{tabular}{| p{7cm} |  p{8cm}  |}
\hline
Category & Sub-Category \\ \hline \hline
\multirow{3}{7cm}{ Modifications to current association studies techniques} & Quality control of the analyzed data \\ \cline {2-2}
&Significance evaluation\\ \cline {2-2}
&Association testing approach\\ \hline \hline
\multirow{4}{7cm}{ Integrative approaches for association studies} &  Integrating several association studies for meta-analysis\\\cline {2-2}
 &  Integrating interaction and functional information with genetic data\\\cline {2-2}
 & Integrating gene expression data with genetic data \\ \cline {2-2}
 & Complementing association studies with pathway information\\
\hline
\end{tabular}
\caption{Different categories in addressing shortcomings in disease association studies}
\label{tab:cate}
\end{table}

 This report surveys  the literature that addresses the shortcomings of current methods the identify genetic disease associations. In addition, it reviews the suggested solutions that either  enhance some aspect of the methodologies, or complement them.  The surveyed topics cover two categories of techniques, as summarized in table~\ref{tab:cate}. The first category, \textit{Modification to current association studies techniques}, comprises techniques that aim to pre-process association data, and the methods used for testing/evaluation of disease associations. The second category, \textit{integrative approaches for association studies}, consists of techniques that aim to integrate additional sources of information into association studies such as functional information, gene expression and pathway data.

The rest of this report is organized as follows. Chapter 2 introduces the history of disease association studies and presents their different types. The  standard statistical approaches used in association studies are also explained in this chapter. Chapter 3 discusses  four of the most prominent challenges faced in identifying genetic disease associations. Chapter 4 describes several techniques that aim to improve different aspects of disease association methods. Chapter 5 discusses various techniques for integrating additional information into association studies. Finally, Chapter 6 summarizes the survey and provides concluding remarks about possible research directions.

\chapter{Disease Association Studies}
The idea of disease association studies originated from the work of  the eugenicist statistician Ronald Fisher on a mathematical model of genetic contributions to complex phenotypes \cite{Wallace2009}.  Based on Mendelian rules, Fisher demonstrated that continuous traits (such as weight and height) and susceptibility to disease  stem from the joint contribution of different genes \cite{Fisher1918}.  This model has led to extensive research in genetic screening in order to identify genetic risk factors in predisposition to different types of diseases. Complex diseases are known to be affected by interacting genetic and environmental factors \cite{Wilkie2001, McCarthy2008}. Therefore, the discovery of genetic causality and mechanisms of actions through which disease develops requires a thorough analysis of genetic data and educated interpretation of disease association studies.

The interest in association studies is justified by the importance of their findings. As mentioned in Chapter 1, the ultimate goals of association studies are prevention of disease, early diagnosis, improved prognosis and the discovery of new therapeutic targets \cite{Khoury2009}. 

\section{Disease Association Studies Data}
 Genetic markers data, i.e. specific parts of DNA such as genes with known properties and locations \cite{Lawrence2005}, is the type of data analyzed in genetic association studies. The analysis of such genetic marker data in disease association studies leads to the detection of associations between common DNA variants and the disease or trait  being studied \cite{McCarthy2008b}. Different types of genetic markers such as Copy Number Variations (CNV) and SNP are used in disease association studies. As defined in Chapter 1, SNPs are single nucleotide mutations. Single nucleotides are the building blocks of the DNA sequences, where a nucleotide is represented as one of four values \{A, T, C, G\}.  CNVs are modifications in the number of nucleotides at certain loci on the DNA, they can be either duplications or deletions of regions of nucleotides \cite{Feuk2006}. Both kinds of mutations (i.e. SNP and CNVs) are caused by several heritable and environmental factors. In the following sections, the term SNPs is used interchangeably with ``markers" or ``variants"  since SNPs are the most common type of variants used in association studies. 

\begin{figure}[tbp!]
 \centering
   \includegraphics[scale=.7]{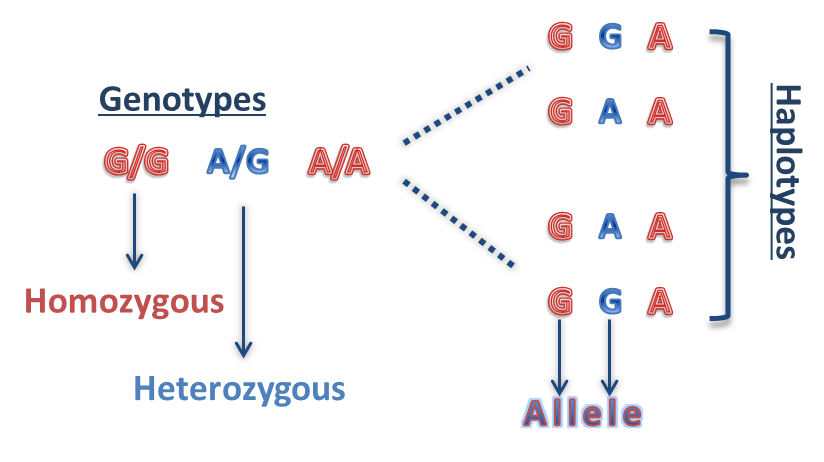}
  \caption{An example of a genotype sequence for three SNPs along with their possible haplotypes. }
\label {fig:gen-hap}
\end{figure}

The majority of association studies are performed using bi-allelic variants where each variant has two alleles, major and minor. A major allele appears more frequently in a population whereas a minor allele appears less frequently in the same population.  The genetic data can be represented as  genotypes or haplotypes. Genotypes are the succession of pairs of alleles for each location (locus) on the DNA  \cite{Lawrence2005}. A genotype is called homozygous when its pair of alleles has identical values (e.g. AA or TT). However when the pair of alleles have different values (e.g. AC or GA), the genotype is called heterozygous. Haplotypes are sequences of statistically related single allele values at successive locations on the DNA \cite{ Brookes2001, Zaitlen2005}. Figure~\ref{fig:gen-hap} presents an example of a sequence of genotypes and its derived haplotypes. It also highlights the difference between homozygous and heterozygous genotypes and demonstrates alleles on haplotypes.

\section{History of Disease Association Studies}
\begin{figure}[tbp!]
 \centering
   \includegraphics[scale=.5]{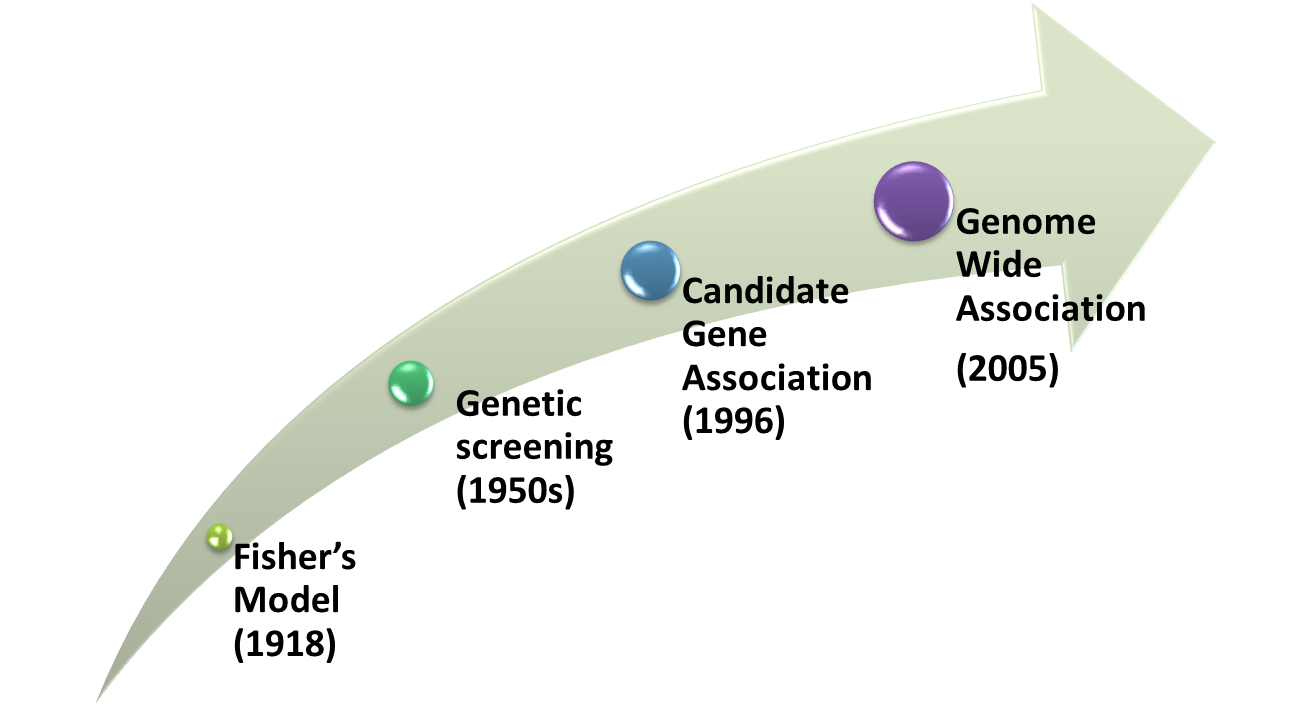}
  \caption{Association studies time line. The time line shows the beginning of association studies with Fisher's model , which led to genetic screening. Genetic screening then paved the way for candidate gene association studies and later to GWAS}
\label {fig:time}
\end{figure}

Following Fisher's model of disease susceptibility, genetic screening started to gain popularity in 1950s where the genetic contribution in predisposition to disease was sought \cite{Wallace2009}. The tobacco industry funded one of the earliest research projects related to the predisposition to lung cancer in 1954 \cite{Wallace2009}. Since 1990s, disease susceptibility research evolved into candidate gene association studies where genetic variants on candidate genes were analyzed  \cite{Risch1996}. In candidate gene association studies,  researchers identify possible target genes of a disease using knowledge about disease pathways and sequence these genes for analysis in association studies. The cost and scarcity of sequencing resources in the past were the reasons for bounding the search to candidate genes. Due to the incomplete understanding of disease mechanisms,the process of acquiring pathway information and then identifying target genes is difficult.  Yet with the advent of recent biotechnologies, sequencing the whole genome became a reality in 2001 \cite{Venter2001} and the first genome-wide association study (GWAS) was performed in 2005 to investigate the genetic risk factors of age-related macular degeneration \cite{Klein2005}. Since then, GWAS is considered the de facto standard in association studies.  Searching for disease associations on a genome-wide scale requires more computational resources than searching on a candidate gene scale . At the same time, the size and coverage of GWAS data increase the probability of encountering genetic variants in association with disease. The evolution of association studies is depicted in Figure~\ref{fig:time}, starting from the Fisher's model in 1918 until the first genome-wide association study in 2005 \cite{Fisher1918, Klein2005}.

\section{Categorization of Disease Association Studies}

\begin{figure}[t!]
 \centering
   \includegraphics[scale=.7]{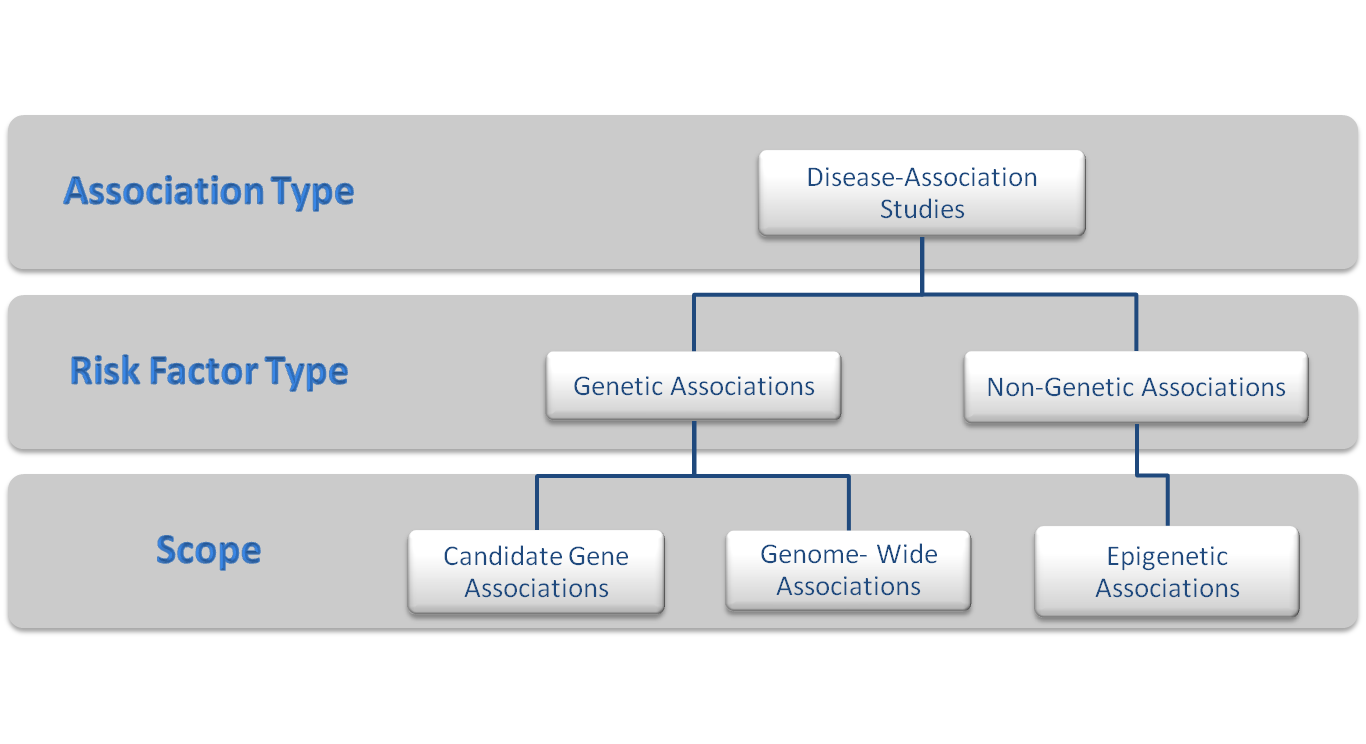}
  \caption{The categorization of association studies. Disease association studies comprise two categories: genetic and non-genetic (such as environmental). Genetic associations can be performed either on a candidate gene or on a genome-wide scope, whereas non-genetic association is performed by analyzing epigenetics data. }
\label {fig:categ}
\end{figure}

Disease association studies can be categorized based on two criteria: \textit{risk factor type}; \textit{scope of the studies}. Risk factors are either genetic or non-genetic. For genetic association studies, the scope can be either limited to the analysis of variants on candidate genes (candidate gene scope) or broadened to cover all variants on the genome (genome-wide scope). The scope of non-genetic association studies is limited to epigenetic association studies \cite{Rakyan2011}. Epigenetics refers to the processes that modify gene functions without causing a mutation in the DNA sequence \cite{Lawrence2005}. Review of epigenetics studies is beyond the scope of this report. Figure ~\ref{fig:categ} shows a graphical representation of the categories of disease association studies. The discussion in the following sections is limited to genetic association studies on a genome-wide level, known as GWAS.

The data analyzed in association studies vary. Some studies are based on the comparison of SNP frequencies between a predefined number of individuals manifesting the disease (referred to as cases) or the trait being studied and a number of individuals who do not manifest the trait (referred to as controls). This type of studies is known as case/control studies. When the case/control data is collected from related individuals in a family, the association study is referred to as a family-based association study \cite{Clarke2011}. When case/control data is collected from unrelated individuals, the association study is known as population-based association study \cite{Clarke2011}.

\section{Standard Statistical Approaches for Disease Association Identification}
Case/control association studies explore the difference in allelic (haplotype data) or genotypic (genotype data) frequencies of a SNP between cases and controls. For instance, if a certain allele has a higher frequency in cases than in controls, the difference in frequency suggests an association of the disease with this specific allele \cite{Lewis2012}.

The standard approach, known as single-marker-based association testing, consists of  inspecting the association of a certain disease or trait with a single SNP at a time \cite{Wang2010}. Another approach, multi-marker-based association testing, is also used to seek patterns in allelic or genotypic frequencies by analyzing groups of SNPs together. This latter way of testing increases the ability of association studies to discover disease risk factors due to Linkage disequilibrium (LD) patterns that provide a mean to find associations with non-assayed variants \cite{Lewis2012, Bakker2005} 

A variety of statistical tests is used to infer association between a disease and a SNP. The allelic or genotypic frequencies of a SNP are compared in order to reject or accept the null hypothesis, that is, the absence of association between the studied disease and the SNP being inspected. The frequencies are summarized in several ways using contingency tables. For allelic frequency comparison, the contingency table contains the counts of major and minor alleles in cases and controls \cite{Clarke2011}. For genotypic frequency comparison, the counts of major/minor homozygous and heterozygous genotypes of cases and controls are summarized in the contingency table \cite{Clarke2011}. Disease penetrance is the probability of a disease to be manifested in individuals who carry certain alleles \cite{Urbach2012}. Additive, dominant and recessive models are examples of disease penetrance  models used in association studies \cite{Ohashi2003}. The presentation of allelic/genotypic information in a contingency table depends on the model of disease penetrance adopted in the study. For example, color blindness or Daltonism is a disease with a recessive penetrance model, which means that a person has to carry two minor alleles in order to manifest the disease. An example of a genotypic contingency table for Daltonism is presented in Table~\ref{tab:cont} where `A' is the major allele and `T' is the minor allele. Due to the recessive penetrance model of Daltonism, major homozygous and heterozygous genotypes do not manifest the disease unlike minor homozygous genotypes.

Using contingency tables, the independence of alleles/genotypes or their trend is tested using one of several statistical tests. If as a result of statistical test an allele or genotype is associated with cases or controls, the null hypothesis is rejected and an association between this allele and the disease being studied is inferred. Consequently, an individual is believed to have a risk factor for a disease if this individual carries the identified allele associated with disease manifestation. 

\begin{table}[tbp!]
 \centering
\begin{tabular}{|l l l | l|}
  \hline
\multicolumn{4}{|l|}{\bf{Genotype counts:}} \\
\hline
 & \underline{AA + AT}&\underline{TT} & \underline{Total}\\ 
\underline{Cases} & \bf{190}&\bf{60} & \bf {250}\\ 
\underline{Controls}& \bf{200}&\bf{25} & \bf{225}\\ 
\hline
\underline{Total} & \bf{390} &\bf{85} & \bf{475}\\
\hline
\end{tabular}
  \caption{An example of an genotypic contingency table for a recessive penetrance disease (e.g. Daltonism) with major allele `A' and minor allele `T' in a dataset of 250 cases and 225 controls. Major homozygous or heterozygous genotypes do not manifest the disease whereas a minor homozygous genotype manifests it.}
\label {tab:cont}
\end{table}

Several statistical tests are used to infer disease risk factors. These tests are categorized into three main classes: \textit{goodness-of-fit, likelihood-based} and \textit{regression-based}. One of the most frequently used tests is the Chi-squared ($\chi^2$), a \textit{goodness-of-fit} test; it infers the presence of an association or its absence by assessing the dependence or independence of rows and columns in a contingency table \cite{Clarke2011, Turner2000}. In order to use  the $\chi^2$ test, cases and controls are presented as categorical data but do not need to be ordered according to major/minor homozygous or heterozygous. This test cannot be used when the disease penetrance model follows a trend in the variations of allele/genotype frequencies (i.e. when the disease risk increases with the number of a certain allele in a given genotype such as `T'). For example, in a certain genotype (GG or GT or TT), if the disease risk increases from none in the `GG' genotype, since there is no `T' allele, to high risk in the `TT' since there is two `T' alleles. Thus, the variation of allele frequencies is believed to have a certain trend that is associated with the disease.  Hence, the disease penetrance model has an impact on the results of the $\chi^2$ test \cite{Clarke2011}.

Another commonly used test is the Cochran-Armitage (CA) trend test. It tests whether the variation in the number of a certain allele, between different ordered categories in the contingency table, follows an upward or downward trend in cases versus controls \cite{Armitage1955, Clarke2011}. Unlike the $\chi^2$ test, the CA is applied to genotype data that is ordered according to one of the disease penetrance models \cite{Clarke2011}. For instance, the above example of Daltonism has a recessive penetrance model, therefore the order of genotypes in a CA contingency table should reflect the number of recessive alleles in each genotype (i.e. the order of genotypes should be $ GG \rightarrow GT \rightarrow TT $).  CA is more robust than $\chi^2$ when the dependencies between the analyzed genotypes is due to the characteristics of the sampling population \cite{Lewis2012}. In general, the $\chi^2$ and the CA tests are used to test for associations when adopting a single-marker-based testing.

Associations can also be inferred using \textit{likelihood ratio} tests.  This type of testing was initially designed for association testing in family-based data \cite{Lewis2012}, but it was extended to support case/control data \cite{Fallin2001}. The advantage of likelihood ratio tests is their applicability in multi-marker-based testing. In these tests, an association is based on the likelihood of having specific haplotypes (i.e. a group of SNPs with specific allele values) in cases versus controls \cite{Clarke2011, Morris1997}. The likelihood of different haplotypes in cases and controls is usually estimated using maximum likelihood estimation.   

Logistic regression, a \textit{regression-based} method, is another approach for association testing yet a more computationally expensive one \cite{Wu2009,Hunter2007, Wellek2012}.   In this approach, SNPs (predictors) are used to predict the disease status (case or control) using several disease penetrance models \cite{Hunter2007}. Due to its complexity, this method is predominantly used with single-marker-based testing. Applying  logistic regression as a multi-marker-based testing is made more feasible when the relationship between SNPs and the disease is summarized in a model and used in the application of the regression  \cite{Wu2009}.  

All the above mentioned statistical methods for association studies are summarized and compared in table ~\ref{tab:stat}.  The comparison is based on the type of the statistical test used, the applicability of a method to single or to multiple markers, the type of data needed for the method and other special characteristics of each method.

\begin{table}[!tbp]
\centering
\begin{tabular}{|p{2cm} |p{2cm} | p{2.5cm} | p{3cm} | p{3cm} | }
\hline
\bf{Method }& \bf{Statistical Base} & \bf{Marker Base} & \bf{Data type} & \bf{Characteristics} \\ \hline \hline
\multirow{2}{*}{$\chi^2$ } & \multirow{2}{2cm}{Goodness-of-fit} & \multirow {2}{3cm}{Single-marker} & \multirow {2}{3cm}{Allellic/ Genotypic} & Widely used method \\ \cline{4-5} 
 & & & \multirow {2}{3cm}{Unordered/ Categorical} & Not applicable if penetrance model has a trend \\ \hline \hline
\multirow{2}{*}{CA } & \multirow{2}{2cm}{Linear Regression} & \multirow {2}{3cm}{Single-marker} & \multirow {2}{3cm}{Allellic/ Genotypic} & find trend in penetrance model\\ \cline{4-5} 
 & & & \multirow {2}{3cm}{Ordered/ Categorical} & robust when allelic/genotypic frequencies changes over generations\\ \hline\hline
Likelihood Ratio Test  & Likelihood & Single/multi-marker& \multirow {2}{3cm}{Allellic /Genotypic} & uses maximum likelihood estimation\\ \hline \hline
\multirow{2}{2cm}{Logistic Regression Test} & \multirow{2}{2cm}{Logistic Regression } & \multirow{2}{3cm}{Single/multi-marker} & \multirow{2}{3cm}{Allellic /Genotypic }& high complexity\\ \cline{5-5}
&&&& explores different penetrance models \\\hline
\hline
\end{tabular}
\caption{Comparison of common statistical tests for association studies}
\label{tab:stat}
\end{table}

\chapter{Challenges in Disease Association Studies}
Despite the success in discovering numerous markers associated with a wide range of diseases, associations studies face several challenges. Knowing and addressing these challenges pave the way for more accurate and novel discoveries of disease risk factors in the future. The challenges reviewed in this report are problems in current methods of discovering disease associations that are frequently discussed in the literature. They span a variety of topics,  from the quality of the data to the details of obtaining and replicating the results. 

\section{Missing Heritability}
The genetic variants identified by analyzing GWAS data can not currently explain a significant amount of the genetic variance present in complex diseases \cite{So2011}.   The total portion of disease phenotypic variations attributed to additive genetic factors is known as the heritability of this disease \cite{Lee2011}.  The \lq\lq{}missing heritability\rq\rq{}is the term used to refer to the unexplained portion of  the genetic variance in diseases \cite{McCarthy2008, Manolio2009}. In literature this is attributed to: 
\begin{enumerate}
\item[i-] the lack of statistical power to detect associations of genetic markers explaining a smaller portion of disease risk than the currently reported ones \cite{Cusanovich2012}.
\item[ii-] rare variants that are not being captured by sequencing technologies. Current bio-technologies enable  the sequencing of millions of common variants on the DNA. 
\item[iii-] the influence of gene-environment interactions, epistasis (i.e. the effect of several genes on another single gene) and epigenitcs \cite{Cusanovich2012, Manolio2009}.
\item[iv-] other forms of genetic variants, such as copy number variations, and the lack of their analysis \cite{Manolio2009}.
\end{enumerate}
In general, there is no consensus on the reasons for the failure to capture  a significant portion of the genetic variation  responsible for individual differences in complex disease susceptibility \cite{Manolio2009}.   

\section{Shortcomings of Traditional Statistical Testing}
The use of some traditional parametric statistical approaches hinders the process of modeling the complexity of genetic architecture of diseases \cite{Moore2010}. Such methods overlook the effect of different types of interactions such as gene-gene interactions on disease, due to their limited ability to uncover high-order non-linear interactions \cite{Moore2010, Cantor2010}. The traditional approach of comparing the distributions of genetic markers in case/control studies uses hypothesis testing and \textit{p-values} to make statistical inferences \cite{McCarthy2008b}. However, this approach does not express the statistical power of the study, the false discovery rate (the percentage of reported significant associations that are truly unassociated with the disease), or the number of likely true positives \cite{McCarthy2008b, Storey2003, Benjamini1995}. This inference method sets a single threshold on \textit{p-values} in order to report an association as significant. The more stringent the threshold the more significant is the association. Given the large number of assayed variants, multiple use of statistical testing, that is, multiple comparisons in association studies leads to a large number of false positives in the reported results \cite{Benjamini1995}. In addition, stringent \textit{p-value} thresholds prevent discovering many causal variants with modest effect  sizes \cite{Gibson2010}. Hence, translating current discoveries in association studies to clinical knowledge is considered a problematic task due to the number of false positives.

\section{Inter-Population Applicability}
In the past decade, the majority of association studies were performed on populations of European ancestry \cite{McCarthy2008, Rosenberg2010}. The ubiquity of technological resources and research institutions and their concentration in some parts of the world as compared to others is one of the  reasons for the focus on European population.  Additionally, some population specific characteristic such as allele-frequency variations has also favored the focus on association studies in European populations \cite{Rosenberg2010}. The findings of studies in one population may not be applicable to other populations \cite{McCarthy2008}. Thus, an association found in one population with a specific effect size does not necessarily translate into the same association in another population. The absence of association across populations can be attributed to environmental and epigentic factors. It can also be due to the difference in effect sizes and external influences effecting different populations. The lack of applicability across populations is yet to be explored and investigated. The importance of tackling this problem stems from the need to understand the effect of reported associations on diverse populations in order to provide improved universal health care \cite{Rosenberg2010}.

\section{Replication Issues}
An essential step after performing an association study is the replication of results in order to avoid spurious associations \cite{McCarthy2008b, Khoury2009, Moore2010}. Replication is considered a validation of the discovered associations and an evaluation in additional independent groups of cases and controls \cite{McCarthy2008b}. As mentioned in Section 3.2, multiple statistical tests performed in association studies lead to a significant number of false-positives, which are highly manifested with the massive amount of data in GWAS \cite{Pearson2008}. The importance of replication arises from the need to have reproducible associations \cite{McCarthy2008b}. However, replication efforts face some difficulties in reaching the desired outcome. Failure to replicate association results can be attributed to several factors. Poor design of an association study, such as the selection of non-population-based controls or errors in genotyping, may lead to the discovery of spurious associations. Heterogeneity of the studied population, i.e. difference, between the original and replication studies is another cause of replication failure \cite{McCarthy2008b,Khoury2009}. Heterogeneity in the studied populations can be caused by sampling biases or difference in allelic frequency within/among populations \cite{Khoury2009}. Heterogeneity may also stem from epidemiological reasons, where some genetic variations are affected by environmental factors \cite{Khoury2009}. Heterogeneity may affect the results of replication in two different ways. First, it may lead to spurious associations, which are reported associations but are not truly associated.  Second, it may affect the estimated magnitude or extent of association as reported in the discovery and replication studies \cite{Khoury2009}. 

\begin{figure}[tb!]
 \centering
   \includegraphics[scale=.7]{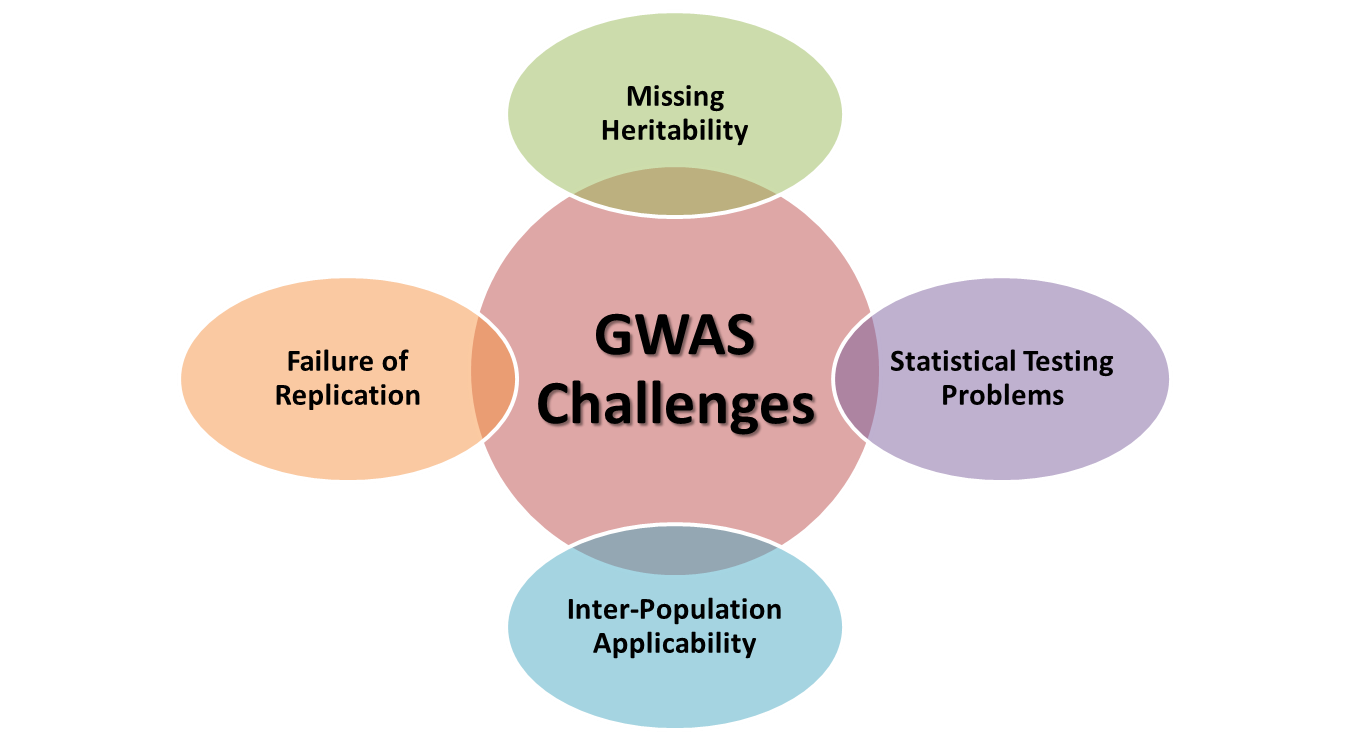}
  \caption{Summary of challenges in association studies (precisely GWAS). Missing heritability, statistical testing problems, inter-population applicability and failure of replication are the challenges discussed in this report.}
\label {fig:short}
\end{figure}

The aforementioned challenges are summarized in Figure ~\ref{fig:short}. The following chapters lists emerging techniques that aim to overcome several aspects of the discussed shortcomings, in particular those of the analysis of GWAS, and to explain additional factors involved in disease. The discussion in the rest of the report focuses on GWA studies rather than on the general concept of association studies.

\chapter{Modifications to Current techniques in Disease Association Studies}
In this chapter, we survey the literature pertaining to modifications to the techniques currently used in GWAS. Such modifications include  either modifying some steps in the process of  association discovery or increasing the genetic coverage and discarding of unwanted structure/ relatedness in the GWAS data. 

The discussed techniques change the standard way of performing GWAS and are categorized into three groups:
\begin{itemize}
\item Methods increasing the coverage of GWAS data by either using imputation techniques (which will defined later in Section~\ref{data_quality}) or aggregating several datasets. In addition, this category includes guidelines and recommendations about the GWA study design that aim to minimize the bias that usually arise in GWA studies.
\item Preprocessing methods for SNP selection designed to assist with choosing the most informative group of SNPs to be included in the GWAS. 
\item Methods that focus on significance evaluation of statistical testing for associations. In this category, we summarize the  modifications to the standard parametric statistical approaches.
\item Statistical and computational methods that aim to improve the association discovery technique itself.   
\end{itemize}

\section{GWAS Data} \label{data_quality}
Due to the advancements in sequencing technologies and imputation techniques, massive amount of data are available for GWA studies \cite{Wallace2009, Li2009}. Imputation techniques are statistical or computational methods used to  predict the values of non-assayed variants depending on the phenomenon of LD, which is the existence of dependencies between SNPs residing in close proximity.  Consequently imputation techniques increase the genetic coverage by increasing the number of SNPs included in the association study \cite{McCarthy2008b}. At the same time, increasing the amount of available data results in higher statistical power of GWA studies. Selecting the group of cases and controls to be included in a study is a critical process. This selection is influenced by many factors such as the population used and the way of ascertaining that a case really manifests the disease or that a  control is really free of the disease.  In some circumstances when the case/control selection introduces a bias towards cases or controls, increasing the amount of data provides a way to improve the accuracy of an association study \cite{McCarthy2008b}.  Increasing both the number of individuals involved in the GWA study and the genetic coverage (number of assayed SNPs) are two important factors that increase the power of an association study \cite{Manolio2009a}. By ensuring a wider genetic coverage, GWA studies are more likely to discover various causal variants with different effect sizes. However, an increased genetic coverage does not lead to the discovery of variants with smaller effects sizes since their discovery depends on the method used to find the disease associations. 

Even though in GWA studies, the number of assayed SNPs may reach 100,000 to 1,000,000,  this coverage is still incomplete since the number of genetic variants exceeds 10 million \cite{Li2009}. Marchini \textit{et al.} \cite{Marchini2007} proposed to impute non-assayed data using a population genetic model and tested the influence of imputed data on the power of the association study. Li \textit{et al.} \cite{Li2009} emphasized the importance of such imputation techniques since they enable the investigation of diseases associations with non-assayed genetic variants. Yet, the authors commented about imputation techniques by anticipating that in the era of genome-wide sequencing, imputation techniques will be mainly used for combining genetic data from different resources for analysis in the framework of meta-analysis of GWA studies \cite{Li2009}. 

In the following subsections, we discuss different GWA data related topics: quality control for  GWAS Data, design recommendations and aggregation of GWA data from multiple sources.

 \subsection{Quality Control and Design Recommendations for GWAS}

The design of a GWA study is a critical step in the process of discovering disease risk factors and avoiding spurious disease associations. It mainly affects the quality of the data in terms of robustness in the face of different sources of bias, for example population stratification or cases misclassification \cite{Pearson2008, McCarthy2008b}. Population stratification is the difference in allele frequencies between cases and controls caused by systematic ancestry differences and not as a result of a disease \cite{Price2006}. The discovery of spurious associations may be the result of population stratification. Therefore, special attention must be paid to the selection and classification  of participants in a study (i.e. cases vs. controls). Controls should be drawn from the same population as the cases and the classification of the population into cases vs. controls should avoid selecting individuals whose disease status is not clearly defined in order to avoid spurious associations \cite{Pearson2008}. Moreover, choosing to sequence variants that belong to genomic areas with lower information redundancy leads to a larger genetic coverage, which in turn increases the power of the study \cite{McCarthy2008b}.

Price \textit{et al.} \cite{Price2006} proposed the use of principal component analysis (PCA) to correct for population stratification \cite{Price2006}. Adjusting the genotype data by removing possible ancestry correlation is required to avoid spurious associations caused by population stratification. The authors  extracted the principal components of a matrix representation of genotype data in order to reveal the directions of variation in the genetic data, which are ancestry-based variations \cite{Price2006}. Then, all genotype data are adjusted by using the Eigen values of the principal components, which represent the directions of variation in the data \cite{Price2006}. 

The quality of the assayed genetic variants used in GWAS also influences the power of the study \cite{McCarthy2008b, Li2009}. Sequencing and genotype calling techniques differ in the quality of their resulting sequences. The assayed genetic variants with poor quality genotypes (reflected by the genotype calling rates) should be filtered out of GWA data \cite{Bravo2010}. Additionally, variants demonstrating a significant deviation from the Hardy-Weinberg equilibrium, a property of variants that reflects how genotype frequencies defer from generation to another in a population, should also be excluded from the GWA data \cite{McCarthy2008b, Li2009}. The above procedures ensure a better quality of the assayed variants, which helps avoid spurious associations \cite{McCarthy2008b, Li2009}. 

\subsection{Consolidation of GWAS Data}

With the purpose of increasing the amount of genetic data available for GWA studies, a consortium of several institutions and foundations such as the National Human Genome Research Institute (NHGRI) called for consolidating the research efforts in GWA studies \cite{Manolio2009a}. They provide the means to combine data from multiple resources and make the integrated dataset available, thus supporting more accurate analysis and more powerful GWAS \cite{Manolio2009a}.  The consortium also introduces some improvement to the ascertainment of cases/controls by adopting strict ascertainment schemes for each disease being studied \cite{Burton2007}. In addition, it emphasizes the importance of using a single control group for the analysis of different diseases as shown in the GWAS of seven common diseases \cite{Burton2007}, unlike previous GWAS where different control groups were used for each disease \cite{ Pearson2008, Manolio2009a}. Another project, 1000 Genomes project, aims to sequence the whole genome of thousands of individuals and make this data available for researchers \cite{Via2010}. This project also facilitates the identification of rare genetic variants using improved imputation techniques to estimate the values of missing variants. Thus, the power of association studies is expected to increase with the availability of such high resolution genetic data \cite{Via2010}.

\section{Preprocessing of GWAS Data for Dimension Reduction} 

SNP selection is the process of choosing the most informative variants to be included in a GWA study. The informative variants, known as tag SNPs, are a group of variants that summarize the information conveyed by the whole set of SNPs. The selection process is performed by informatively decreasing the dimension of the SNP dataset to facilitate association testing. This selection does not defeat the purpose of increasing the coverage of the GWAS dataset since the group of selected SNPs convey the information hidden in the original set of SNPs but without redundancy. Selecting a subset of SNPs to be assayed and analyzed is considered a way of saving sequencing resources and sometimes providing a higher quality data \cite{Li2009, Davidovich2009}. 

In the literature, different categories of approaches are proposed for SNP selection. Dimension reduction methods borrowed from the data mining literature as well as machine learning techniques are used for SNP selection \cite{Moore2010}. SNP selection can  be performed depending on the intrinsic properties of the variants or their prediction accuracy.  Keating \textit{et al.} used the correlations among different groups of SNPs due to LD, an example of SNP properties, as a criteria to select a representative variant for each group of correlated  SNPs \cite{Keating2008}. Miclaus \textit{et al.} used the deviation from the Hardy-Weinberg disequilibrium, another example of SNP properties, for the selection of the most infromative set of variants \cite{Miclaus2009}.  As for prediction accuracy, it can be used for SNP selection in two ways: 
\begin{itemize}
\item Select the set of variants that can accurately predict  the value of other SNPs in the data. The discussion in this section is limited to this type of SNP selection techniques.
\item Select the set of variants that can best classify the data in the desired disease outcome.
\end{itemize}

  SNP selection techniques that optimize prediction accuracy   are used to decrease the number of variants to smaller subset with minimal redundancy in the information. Prediction techniques provide a mean for selecting the most informative SNPs according to how accurate a select group of SNPs can predict the values of non-selected group \cite{Davidovich2009}.  Chuang \textit{et al.} \cite{Chuang2009} adopted a Genetic Algorithm (GA) with the predication accuracy of k-nearest neighbor (KNN) as a fitness function in order to select the most informative SNPs.
   Davidovich \textit{et al.} \cite{Davidovich2009} demonstrated that selecting SNPs for an association study using the prediction accuracy also increases the accuracy of the study. The main idea behind  Davidovich \textit{et al.}'s  work is to compare two methods of SNP selection: the first uses SNP correlation as a selection criteria and the second uses prediction accuracy, in order to evaluate how the power of association is affected by either selection criteria \cite{Davidovich2009}.

\section{Significance Evaluation }

Reaching an accepted significance level in a genome-wide context is at the core of GWA studies \cite{Ioannidis2009}. The standard approach to ensure genome-wide significance is setting a \textit{p-value} threshold of  less than or equal to $10^{-7}$ \cite{Ioannidis2009}. Lower \textit{p-value} threshold such as the stricter $10^{-10}$ \cite{Ioannidis2009} may result in more significant associations, yet it introduces a lot of stringency and may lead to missing associations with smaller effect sizes and contribute to the missing heritability \cite{McCarthy2008b}. Genome wide association studies require a large number of statistical testing, thus a correction of \textit{p-value} thresholds is needed to avoid an increased number of false positives. The most widely used correction is the Bonferroni correction where the \textit{p-value} threshold is divided by the number of performed tests \cite{Pearson2008, Benjamini1995}. However, Pe'er \textit{et al.} \cite{Peer2008} raised the point that the use of the simple Bonferroni correction overlooks the fact that variants associated with the disease being studied are not all independent.  

In addition to the fact that many reported associations do not meet strict \textit{p-value} thresholds, using only \textit{p-values} to report association significance does not provide all the information needed about a study \cite{Nguyen2009, Berger1987}.  The statistical power of the association study as well as the confidence in the reported associations greatly depend on the amount of the data, which is not conveyed in the \textit{p-value}. For instance, a \textit{p-value} of an association between a disease and a certain variant may have the same value in two different studies where the sample sizes, the effect sizes, and the segregation into cases and controls are greatly different \cite{Nguyen2009}. However, the evidence of association in both of these studies may not be equal \cite{Nguyen2009}.  Hence,  when using \textit{p-values} alone, the quantification of confidence in the reported associations becomes difficult \cite{Stephens2009}.   The number of true positives and the false discovery rate both reflect the significance of reported associations and are also not accounted for by the use of \textit{p-values} \cite{McCarthy2008b, Storey2003}.  

Evaluating the significance of disease associations using other measures than \textit{p-value}s is the basis for several studies \cite{Storey2003, Stephens2009}.  The following subsections explains the use of the false discovery rate and a bayesian approach as an alternative to the use of \textit{p-values}.

\subsection*{\textit{False Discovery Rate and $\bf{\textit{q-Values}}$}}
In the context of GWAS, False Discovery Rate (FDR) is the proportion of statistically significant discovered associations that are  not truly associated with the studied trait \cite{Storey2003, Pearson2008}. Reaching the smallest number of false positives ($F$) and the largest number of true positives ($T$) is important; FDR is calculated as the ratio of $F$ to the total number of statistically significant variants ($F+T$) as in Equation~\ref{eq:fdr} \cite{Storey2003}.
\begin{equation}
FDR = \frac{\#~of~False~Positive~Variants}{\#~of~Statistically~Significant~Variants} = \frac{F} {F+T}
\label{eq:fdr}
\end{equation}

The false positive rate is the probability of reporting the variants  that are not associated with the trait being studied, as significantly associated with these traits \cite{Pearson2008}. As mentioned above, the use of false positives rate as the only significance measure overlooks several aspects of the assessment of significant associations. Tenesa \textit{et al.} \cite{Tenesa2008} used the false discovery rate as an assessment of their reported risk loci in association with colorectal cancer. Storey \textit{et al.} \cite{Storey2003} proposed the use of \textit{q-values} instead of \textit{p-values} as a measure of significance. The \textit{q-value} is an FDR-based measure that provides more information than a \textit{p-value} about significant associations and facilitates further analysis \cite{Storey2003}. The \textit{q-value} of a certain variant is defined as the proportion of false positives one gets when calling this variant statistically  significant\cite{Storey2003}.  Whereas the \textit{p-value} is the probability of a non-associated variant to be as or more extreme than the observed probability, the \textit{q-value} is the expected proportion of false positives among all variants to be as or more extreme than the observed proportion \cite{Storey2003}.

\subsection*{\textit{Bayesian Approach}}\label{Bayes}

Bayesian approaches were developed for association studies in order to alleviate the problems of \textit{p-values} using Bayes' rule\cite{Stephens2009}. In Bayesian approaches, the ratio between prior and posterior information about two compared data models is used to measure the strength of evidence favoring one of these models \cite{McCarthy2008b}. Bayesian measure of evidence can be directly compare associations within or across studies \cite{Stephens2009}. Such approaches allow and justify the incorporation biological information in the GWAS analysis using a quantitative way \cite{Stephens2009}.  While using the Bayesian approach in GWAS, the evidence of association and its significance is measured using a quotient known as \textit{Bayes Factor} (BF), as shown in Equation~\ref{eq:bf} \cite{McCarthy2008b, Stephens2009}. The larger the BF the greater the evidence of association. The BF is  the ratio between the probability of the variants to be associated with the disease and the probability of the variants to be observed without the presence of the disease being studied \cite{Stephens2009}. The BF is calculated according to the following equation: 
\begin{equation}
BF = \dfrac{P(data\: |\: H_1) }{P(data\:|\: H_0)}~.
\label{eq:bf}
\end{equation}

The distribution of  data under $H_0$ and $H_1$, in Equation~\ref{eq:bf}, strongly depends on the selected genetic model (additive, dominant, recessive, etc ... ). The use of the BF avoids the stringency of a single \textit{p-value} threshold by assigning probabilities to each of the alternate hypotheses and provides a simpler way of comparing evidence of association across studies \cite{McCarthy2008b}. Craddock \textit{et al.} \cite{Craddock2010} used the BF to assess the evidence of association of copy-number variations with eight different diseases. Additionally, Yasuno \textit{et al.} \cite{Yasuno2010} also used BF to evaluate the significance of the loci they reported to be associated with  intracranial aneurysm.

Despite advantages of BF, a common and valid concern arises when deciding to use it as the significance measure. In order to be able to calculate the BF, the parameters of the prior-distribution, the plausible values of our belief about the possibility of a SNP to be associated with a disease, and the disease genetic models are required to be pre-defined. One way of determining these parameters and values is to use previous GWAS results, if there exist such studies \cite{Stephens2009}.

\section{Association Discovery Techniques}
Single-marker-base association testing is the traditional statistical approach in GWA studies. The data is tested, one SNP at a time, and a \textit{p-value} is calculated to decide between accepting the null hypothesis (absence of association) or the alternative hypothesis (presence of association). Despite the success in discovering thousands of disease-associated SNPs using the traditional approach, improved association approaches are required to overcome the challenges faced in GWAS such as the missing heritability. The majority  of association testing approaches in the literature are designed to increase the confidence in the reported associations. Additionally, they aim to eliminate the effect of possible structures (such as individual relatedness) in the data causing inflation in association results or leading to spurious associations. Some of the approaches are based on single-marker methods and provide a way to increase the power of discovery of associations studies, for example, by decreasing the complexity and speeding up the testing process. Others are multi-marker based, where data from a group of SNPs is tested for association discovery. We categorize the different approaches into two main types: statistical and computational.

\subsection{Statistical Modeling}

The statistical approaches surveyed here use linear modeling and its variations such as the mixed linear model and generalized linear model to model the various factors affecting the disease status (also known as the disease outcome).
Le \textit{et al.} \cite{le2011} use a generalized linear model to represent the relationship between the disease status of individuals (represented as a vector $Y$) and the genetic data (reperesented as a matrix $X$). The model assigns a coefficient for each SNP represented as column in $X$ to designate its contribution in the disease status. To test the null hypothesis of no association, the authors use the a statistic test, which tests the influence of two or more samples, here SNPs, on a certain outcome. A separate statistic is calculated for each SNP in the data and the overall statistic is the average of all SNP statistic values. The authors propose a multi-marker method where the effect of  groups of SNPs is tested as opposed to testing the effect of single SNPs only.  Accordingly, a \textit{p-value} is calculated for each SNP and an overall \textit{p-value} for each group of SNPs is then computed. A Bonferroni correction was applied to correct for multiple testing. 

In order to account for different factors in disease risk, Kang \textit{et al.} \cite{Kang2010} proposed modeling the variance of the disease status as a linear mixed  model (LMM).  The contribution of risk factors such as the environment and epistasis are included in the variance model. In addition, they justified their proposal by emphasizing the importance of such models in estimating the unwanted relatedness between individuals. The variance of the disease phenotype $Var(Y)$ is modeled as a linear combination of the variance of the additive genetic effect and the environmental variance weighted by the probability of two variants at certain locus to be identical by decent. The implementation of such model is computationally expensive due to the multiple calculation of variances. Yet, the authors made some assumptions to speed-up the computation of variances in order to facilitate its use with very large genetic data and called their method Efficient Mixed Model Association Expedited (EMMAX) \cite{Kang2010}. With these assumptions come approximations, which may cause a decrease in the statistical power when compared to exact calculations. 

Zhou \textit{et al.} \cite{Zhou2012} proposed a new efficient implementation of linear mixed models as a single-marker-based approach. Their implementation is faster than previous ones and performs exact calculations instead of approximations. In the literature, several approximation methods were adopted to make LMM applicable on a genome-wide scale. The method proposed by Zhou \textit{et al.} \cite{Zhou2012} is called Genome-wide Efficient Mixed Model Association (GEMMA). Their method modifies an existing method known as Efficient Mixed Model association (EMMA) proposed by Kang \textit{et al.} \cite{Kang2008}. To speed up the testing process, GEMMA avoids the iterative eigen decomposition used in EMMA by substituting it by one eigen decomposition and simple vector-matrix multiplication. The vector of disease status is modeled as a linear combination of covariates, the marker genotypes weighted by their effect size and the relatedness information in the data. The association testing method requires the calculation of  the maximum-likelihood estimates of the model parameters. Since calculating the exact estimates is computationally expensive, especially because of the size of the data, the method instead proceeds by optimizing the log-likelihood functions for each of the analyzed SNPs \cite{Zhou2012}. 

A multi-marker association test was proposed by Kwee \textit{et al.} \cite{Kwee2008}.  Despite the multivariate aspect of their proposed statistical test, its advantage relies in its abilities to produce test statistics with a lower amount of calculation than traditional multivariate approaches without sacrificing the power of the study.   The genotype data is modelled using a non-parametric function known as a kernel and its relationship with the disease outcome is represented as a combination of the modelled genotype data and their estimated covariates. The parameters in the proposed semi-parametric model are estimated using a maximum-likelihood approach. 

The above surveyed literature lacks a discussion about the scalability and the required computational time of the used methods. Although in some of the papers, the authors mention the applicability of their method on a genome-wide scale, the information about exact amount of data that the method can analyze and a complexity analysis is missing. With the advancement in biotechnology and the increasing amount of available data, even the term genome-wide scale is supposed to be clearly defined in order to be able to assess the scalability and applicability of the proposed method. For instance, it is required to know how a proposed testing approach will perform with an increase in the amount of SNPs from 500k to 3 million (where both number are usually claimed to have a genome-wide coverage).  Moreover, in some of the surveyed papers, the authors claim that their proposed method is faster/more efficient than a previously devised method. Yet, they overlook the need to clearly describe the extent of improvement their method provides in terms of computational resources (time and space). For instance, GEMMA \cite{Zhou2012} is claimed to be faster than EMMA \cite{Kang2008} but the discussion is missing the information about how much faster it is.

\subsection{Computational Approaches}
Using parametric statistical approaches for association testing entails modeling the relationship between a phenotype and genetic markers under assumptions that are unrealistic most of the time such as assuming independence between the effects of genetic markers despite the important role of epistasis. In addition, parametric statistical approaches do not model high-order non-linear interactions that play an important role in disease risk factors \cite{Moore2010}. Thus, computational approaches such as machine learning and data mining methods are proposed instead \cite{Moore2010}. 

\begin{figure}[!tb]
 \centering
   \includegraphics[scale=.4]{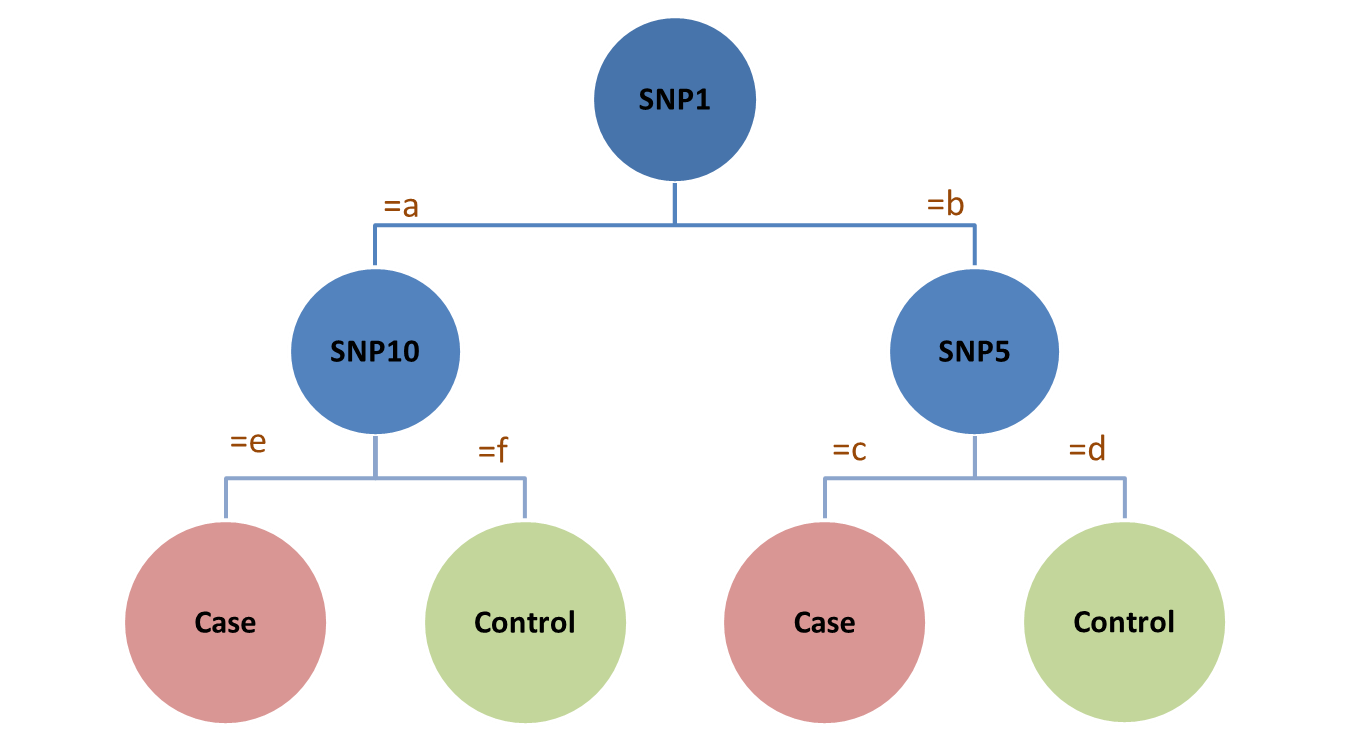}
  \caption{An example of a decision tree where each node represents a SNP and  branching is based on the allele value at each node.The leaves denote the classes: Either \textit{case} or \textit{control}. }
\label {fig:DT}
\end{figure}

Moore \textit{et al.} \cite{Moore2010} discussed the use of decision trees and random forests  for testing associations between a number of SNPs and a disease phenotype. In the data mining literature, these two classifiers are used for distinguishing between classes based on information carried by the attributes (i.e. features and characteristics; here SNPs). A decision tree is a predictive model used to classify an object (here an individual) into a predefined set of classes depending on the values of the object's attributes (SNPs) \cite{Rokach2007}. The genetic model underlying the disease risk is represented as a decision tree or a random forest where the values of associated SNPs are used to predict the disease phenotype. An example decision tree is shown in Figure ~\ref{fig:DT}. Each node in the tree, known as a predictor, represents a SNP for example SNP5 in Figure~\ref{fig:DT}. The nodes at the very last level of the tree, known as leaves,  represent the classification outcome, for example in Figure~\ref{fig:DT} the leaves are either \textit{case} or  \textit{control}. While building the tree, going from one level to another, a process known as branching, is performed based on the possible values the node in the previous level can have. For example to go from level 1 to level 2 in Figure~\ref{fig:DT}, the value of SNP1 can be either ``a" or ``b" and these values lead to node SNP10 or SNP5 respectively in the next level .  The proposed method is a multi-marker-based association testing as the decision tree's classification outcome depends on a set of SNPs rather than on a single one.

The trees are built using the following steps:
\begin{itemize}
\item Select a portion of the data to be used as a training set for building the tree classifier;
\item Randomly choose a subset of SNPs from the training set to be used as predictor of the disease class at each node of the tree;
\item Decide on selecting the nodes to be added at a certain level of the tree and their branching out  by choosing the highest information gain caused by each of the selected SNPs \cite{Kotsiantis2007};
\item Recursively repeat the process of SNP subset and best branching selection until the stopping criteria is satisfied then branch out the nodes in the previous level into leaf nodes that represent the disease phenotype (classification outcome).
\end{itemize} 
One of the main advantages of decision trees is the fact that they are self-explanatory \cite{Rokach2007}. However decision tree classification strongly depends on the existence of features with high effect on the classification outcome. When the classification outcome is affected by many complex interactions, decision trees classification tends to get less accurate \cite{Rokach2007}. 

Random Forests are an ensemble of decision trees constructed by randomizing the branching at each node \cite{Rokach2007}. The generation of such an ensemble aims to increase the accuracy of classification by providing a voting scheme where the best classification accuracy is reported as the random forest accuracy \cite{Rokach2007}. Moore \textit{et al.} generated the random forest after performing a bootstrap sampling of the SNP data \cite{Moore2010}. This method has been applied to genetic data taken from patients suffering from asthma, rheumatoid arthritis, glioblastoma and age-related macular degeneration \cite{Moore2010}. Yet, the authors of this paper overlooked discussing the complexity of the proposed methods and did not mention the amount of data used in the application of the methods.

Li \textit{et al.} \cite{Li2009a} proposed a framework for multi-marker-based
 association testing using Principal component Analysis (PCA). They modeled the effect of each SNP on the phenotype of interest using a linear combination of allele frequencies. The effects of a predefined number of SNPs are aggregated as a score for the gene containing the SNPs. Genotype scores are assigned weights proportional to the linkage disequilibrium (LD)  between the corresponding pairs of SNPs. The information about LD is collected from the International HapMap project \cite{Li2009a}.  In order to further decrease the redundancy in the information, carried by the genotype scores and caused by LD,  PCA is applied to the matrix representation of genotype scores with dimensions $n\times M$ where $n$ is the number of individuals in data and $M$ is the number of SNPs in the reference dataset.  Principal components are independent linear combinations of the gene scores. Each principal component (PC) demonstrates a direction of high variation in gene scores and is calculated according to Equation~\ref{eq:pca}:  
\begin{equation}
 PC_l = \sum\limits_{k=1}^M e_{l,k} S_k ,
\label{eq:pca}
\end{equation}
where $e_l$ is the Eigen vector,   $l = 1,...,m$, $m$ is the number of PCs and $k$ is the index of the group of SNPs such that $1 \leq k \leq M$. Finally using a linear regression analysis the principal components are tested for association with the phenotype of interest (disease) \cite{Li2009a}. The authors applied their proposed approach on a data of  550K SNPs and 2527 individuals, however, they did not discuss the scalability of their proposed approach.

Wang and Elston \cite{Wang2007} also proposed a multilocus association test but instead of PCA it is based on weighted Fourrier transformation coefficients. They transform the numerical representation of sequence $X$ of $m$ SNPs for individual $i$ by means of a discrete Fourier transformation (FT) , which maps the data into its frequency components \cite{Wang2007}.
The authors claim that the lowest-frequency FT component compresses the information and variation relayed by the whole SNP data, thus this FT component can be used to test for associations. To emphasize the importance of the lower frequency FT components, FT components are assigned weights inversely proportional to the order of the component in the FT decomposition. Consequently, the weight scores are used to test the association of the group of SNPs with the disease of interest. In comparison with standard testing methods, the FT based test shows an increased power when associations exist between the disease and the low-frequency components. The proposed method has been applied to simulated datasets that contain small amounts of SNPs (4, 10 and 22 SNPs), the fact that casts doubts on the scalability of the method on a genome-wide scale knowing the computational demands of FT.

The matrix decompositions needed for PCA and FT are computationally expensive and require a significant amount of time to be performed, especially when the dimension of the data is very large. The reviewed literature for computational methods does not discuss the scalability of the proposed approaches. The increasing amount of available data and the need to increase the coverage of the genetic information make scalability a pressing requirement. Hence, the category of computational association testing approaches is a potent resource of improvement for GWAS where the applicability of algorithms capable of analyzing ultra-large scale data in GWAS is yet to researched.

\chapter{Integrative Approaches for Disease Association Studies}
Researchers commonly adopt GWAS  as the first step in identifying culprit genes and revealing disease risk factors. As discussed so far in this report, current GWAS results are often not sufficient to identify the details of genetic and molecular mechanisms pertaining to disease due to the missing heritability. Therefore, several approaches were developed to provide further analysis of the reported results. Some approaches aim to combine the evidence of association with genetic functional and pathway data. Other approaches are designed to uncover hidden low-effect associations caused by epistasis \cite{Cantor2010},  whereas another group of approaches tackle the  discovery of disease risk factors by analyzing non-genetic factors at the molecular level \cite{Rakyan2011}. All these approaches have the potential to increase confidence in reported disease associations and to provide additional information to support the discovery process. 

The approaches discussed in this chapter do not change association testing methodologies. Instead, they take advantage of available biological and medical knowledge to provide additional sources of information for GWAS in order to assist in identifying disease risk factors. Some of these approaches are post-processing of GWAS results used to explain how the reported associations influence the disease and its symptoms and help the clinical interpretation of GWAS \cite{Cantor2010}.

In this chapter, we classify the aforementioned approaches into four categories.  The first aggregates GWAS results in form of meta-analysis. The second encompasses the approaches that couple GWAS with  pathway analysis. The third is the integration of functional information about genes with GWAS. The last category comprises methods that combine gene expression data with GWAS. It is worth mentioning that some recent research efforts are directed toward integrating more than one source of information with GWAS in a single framework \cite{Wang2011, Goh2007, Sieberts2007, Schadt2009}. For instance, Schadt discussed the use of molecular networks to bridge the gap between GWAS findings and clinical medicine \cite{Schadt2009}. These networks are build by integrating different types of data: DNA-variation, gene expression, protein interaction proteomic and metabolic data \cite{Schadt2009}. He described the importance of molecular networks in elucidating complex interactions between genes that affect disease etiology and susceptibility. Schadt presented a probabilistic causal networks as an alternative to interaction networks that are built using only gene expression \cite{Schadt2009}. The former type of networks is capable of integrating different sources of data and providing means to infer the effect of one gene on another and its effect on diseases. Bayesian network construction approaches are used to build probabilistic causal networks. The advantage of the Bayesian approach is the ability to incorporate causal relationships using Bayes rule \cite{Schadt2009}. Following a similar integrative approach of multiple types of data, a framework, called  Integrated Complex Traits Networks (iCTNet), was recently proposed by Wang \textit{et al.} \cite{Wang2011}. iCNet integrates information from different sources of data about:  disease-associations, protein-protein interactions , and relationships between disease-tissue, tissue-gene and drug-gene \cite{Wang2011}.

\section{Meta-Analyses}

Meta-analysis of GWAS aims to aggregate the results of different GWAS, examining the same research hypothesis for a common disease while varying either the number of assayed genetic markers/individuals or the study designs. This type of analysis is designed to boost the accuracy of GWAS and to explore the variability in effect sizes and genetic models across studies \cite{ Bertram2007, Ku2010, Stephens2009, Willer2010}. In other words, meta-analysis studies serve to pool evidence in order to increase the chance of finding more disease associations and to decrease the number of false positives \cite{Cantor2010}. The purpose of performing meta-analysis  differs from one study to another.  Some studies assess previously reported associations and investigate the reasons of failure to replicate results \cite{Lohmueller2003}. Others provide a way to increase the number of discovered causal genetic markers with a moderate to small effect on the disease of interest by combining different sources of genetic data previously analyzed in GWAS \cite{Ku2010}. The combination of genetic data assists in discovering rare variants with large effect on the disease being studied and helps in finding the missing heritability \cite{Lee2011}. To perform a GWAS meta-analysis, the  information obtained from the different studies included in the analysis is summary association statistics about variants in each of the GWAS such as odds ratios or effect sizes \cite{Zeggini2009}. Hence, in the rest of this section, the term studies is used to refer to the information obtained from the studies and included in the meta-analysis.

\begin{figure}[tb!]
 \centering
   \includegraphics[scale=.4]{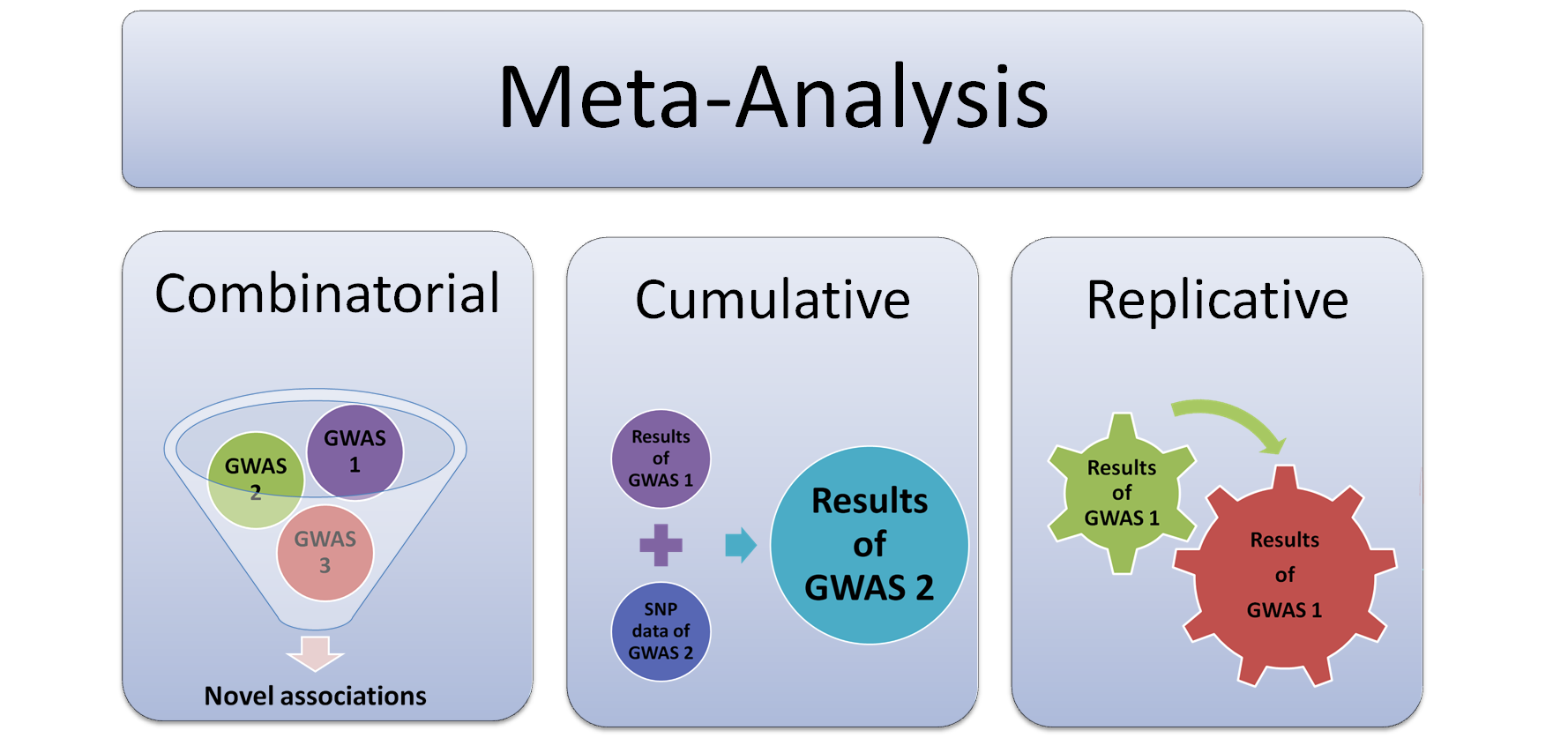}
  \caption{Meta-analysis of GWAS can be performed in different ways. The first consists of combining previous GWAS and using them to discover novel association. The second is the cumulative analysis where the results of previous GWAS are used as prior for another GWAS. The third is the replicative meta-analysis where results from one GWAS are replicated in another data sample. }
\label {fig:meta}
\end{figure}

Cantor \textit{et al.} \cite{Cantor2010} discussed meta-analysis studies as an approach to asses associations across GWAS without the need to re-analyze the original data. Their meta-analysis was designed to prioritize previously reported associations. They differentiated between different types of meta-analysis (as summarized in Fig ~\ref{fig:meta}): 
\begin{itemize}
\item[i)]Cumulative meta-analysis where previous GWAS are used as priors for a new GWAS.
\item[ii)]Combinatorial meta-analysis where previous GWAS are combined together to discover novel associations.
\item[iii)]Replicative meta-analyis where previous GWAS results are replicated using another data sample.
\end{itemize}
\subsubsection{\textit{Combinatorial Meta-Analyses}}
Combinatorial meta-analysis of GWAS can be performed using different methods. The traditional approach is combining \textit{p-values}, however this approach has several limitations including the inability to provide effect sizes, as discussed by Zeggini \textit{et al.} \cite{Zeggini2009}.  The combined studies are weighted using either the sample size or the inverse of estimated variances since studies with low precision (high variance) should have a lower influence on the meta-analysis than the studies with higher precision (low variance) \cite{Zeggini2009}. One way of combining and weighing summary results or datasets of GWAS is using a Bayesian meta-analysis approach \cite{Cantor2010, Stephens2009}. Bayesian hierarchical models are an example of Bayesian meta-analysis where external LD information is used as the priors for studies. These hierarchical models allows for the inclusion of various sources of information, such as results of previous GWAS and SNP functionality as priors \cite{Cantor2010, Stephens2009}. After performing the meta-analysis, SNPs are prioritized either according to meta-analysis \textit{p-values} or Bayes factor (described in Section~\ref{Bayes}). 

\subsubsection{\textit{Cumulative Meta-Analyses}}
In the literature, there are various methods used for cumulative meta-analysis of GWAS.   Using a gene-based approach, Liu \textit{et al.} present a cumulative meta-analysis method based on a Monte-Carlo simulation \cite{Liu2010}. The \textit{p-values} of $n$ SNPs on a gene are obtained from previous GWAS and are used to assess the association of the gene with the disease of interest. The \textit{p-values} of SNPs from different studies are gathered and the genes in which the analyzed SNPs occurred are identified.  Each gene is ranked according to its SNP association \textit{p-values} converted to chi-square statistic.  The proposed Monte-Carlo approach is used to account for the different linkage disequilibrium structures between SNPs on the analyzed genes and it avoids performing computationally expensive permutations to estimate the distribution of these structures under the null hypothesis (absence of association). The authors claim that their proposed method is faster than previous meta-analyses and can be performed without having the individual genotype information from the different studies.

Willer \textit{et al.} proposed two different approaches to perform cumulative meta-analysis and implemented them in a tool, called METAL \cite{Willer2010}. The two approaches are used to combine evidence of association across different studies. The first approach converts the previously calculated \textit{p-values} into z-scores (a score that reflects how many standard deviation a result is away from the mean of a normal distribution), where low values in z-scores reflect low risk associations and high values designate high risk associations \cite{Willer2010}. The second approach uses the calculated standard error for each included GWAS to weigh the effect-size estimates of all reported associations in the study and these weighed values are used as z-scores.     Associations are then compared according to their weighed effect sizes, and overall \textit{p-values} are calculated using the z-scores \cite{Willer2010}. The second approach can only be used when the effect sizes and/or standard errors are given in consistent units across studies whereas the first approach can be used even if the estimates are in different units (for a detailed explanation of when to use either of these approaches see reference \cite{Willer2010}). The authors claim that, asymptotically, both approaches are equivalent when analyzing a trait with the same distribution across the data.

\subsubsection{\textit{Replicative Meta-Analyses}}
Replicative meta-analysis of GWAS can been carried out using different approaches. Lohmueller \textit{et al.} \cite{Lohmueller2003} performed a replicative meta-analysis on 301 published studies of common disease, which cover 25 reported associations in order to discover inconsistency when attempting to reproduce association studies. The analysis led the authors to believe that common variants with modest or low effect play an important role in common disease susceptibility due to their discovered associations. As in all meta-analysis studies, the associations to be included go through a thorough selection process, which depends on their significance and reproducibility in the original study. Association studies with \textit{p-values} lower than a predefined threshold are compared to \textit{p-values} of similar studies generated by simulation. A $\chi^2$ goodness-of-fit test is used to test each association\rq{}s homogeneity of results across all studies. In case of persistently heterogeneous associations between the analyzed studies, heterogeneity is attributed to either flaws in the methodology or population stratification. 

Bertram \textit{et al.} \cite{Bertram2007} performed a replicative/cumulative meta analysis of  Alzheimer\rq{}s disease GWAS. In their study, they re-assessed the significance of one of the most reported associations with Alzheimer\rq{}s and discovered an additional 20 associations with modest effect on Alzheimer\rq{}s. The meta-analysis was performed using the estimates of the odds ratios and 95\% confidence intervals obtained from the Alzheimer's disease published studies.

The work surveyed above about meta-analysis focuses on combining the results of several GWAS in different ways in order to find novel associations or replicate previously reported ones. Yet, it overlooks the benefit of integrating other sources of information with the combined GWAS results such as prior knowledge about pathways, epistasis, protein interactions in order to prioritize the previously reported disease-associations. For example, the confidence in a disease association with a variant occurring on a gene that has been reported to be highly expressed in tissues affected by the disease of interest is greater than the confidence in an association that its only evidence is a \textit{p-value} lower than a certain threshold. Therefore, such a prioritization approach provides the means to accumulate evidence about associations beyond the mere fact of being statistically significant.  Associations supported by prior-knowledge will have a higher impact on the meta-analysis, and associations without this support will have less of an impact.

\section{Pathway Analysis}\label{pathway}
Disease risk is commonly believed to be influenced by a plethora of interacting genes, cell pathways and complex molecular networks \cite{Wang2010}.  Biological cell pathways is a network of genes and molecules that interact in order to accomplish the activities required for a specific cell process such as apoptosis (i.e. programmed cell death). Therefore integrating pathways information with GWAS data has much potential for discovering novel disease risk factors and increasing the confidence in previously discovered ones as well as facilitating the clinical interpretation of disease etiology and mechanisms of action. Pathway collections are publicly available in several databases such Kyoto Encyclopedia of Genes and Genome (KEGG), Biocarta and Gene Ontology (GO) \cite{Wang2010}.  Recently, pathway based approaches are used for the analysis of GWAS in either of two ways:
\begin{itemize}
\item[i)] Instead of focusing on the disease associations that are manifested by strong evidence (very low \textit{p-values}), pathway-based approaches looks for disease risk factors among groups of SNPs (mapped to their closest genes) with small evidence of association. These approaches are designed to search for the associations between a disease of interest and a group of genes where variants (SNPs) belonging to these genes are in modest association with the disease. These groups of genes usually belong to the same pathway.
\item[ii)] Association results are post-processed by taking advantage of the publicly available pathway information. When the analysis reveals that a gene, where a variant associated with a disease occurs, is involved in a certain pathway, the clinical interpretation of the symptoms and mechanisms of action of this disease become more comprehensible. In addition the confidence in association results is increased when these results are supported by pathways information since the evidence of association is beyond just statistical significance.
\end{itemize} 

Wang \textit{et al.} proposed the use of a modified Gene Set Enrichment Analysis (GSEA) with GWA data \cite{Wang2007a}. GSEA is a method originally devised for the analysis of microarray gene expressions, it identifies groups of genes belonging to the same pathway that are differentially expressed instead of focusing on single genes with strongest evidence of differential expression (\cite{Subramanian2005, Wang2007a}. Using pathway-based approach with GWA results facilitates the interpretation of  these results for identifying complex-disease risk. The proposed approach calculate the statistic of association and \textit{p-values} for all SNPs in a GWAS data and assigns them to their closest gene while disregarding SNPs 500kb away from any gene. The highest \textit{p-value} of SNPs residing on a gene is assigned to this gene as its statistic value. The SNPs are assigned to gene sets extracted from the following pathway databases: BioCarta, KEGG and GO, where a gene set is a group of SNPs that share a common biochemical pathway or are coexpressed in previously published experiments \cite{Subramanian2005}. All genes in each gene set are ranked according to their \textit{p-value} obtained from their associated SNPs. Afterwards, an Enrichment Score (ES) is calculated to reflect the overrepresentation of a gene or a gene set at the top of the list of ranked \textit{p-values} \cite{Wang2007a, Subramanian2005}. Pathways related to the highly enriched gene sets are  identified at the end of the analysis. These highly enriched gene sets constitutes the discovered disease risk factors that can not be discovered by only looking at SNPs with \textit{p-values} below certain threshold. This proposed approach was applied to Crohn Disease (CD) data by Wang \textit{et al.} \cite{Wang2009} where they identified and replicated the association of genes in different pathways with CD. The identified pathways and their corresponding cellular processes help provide interpretations to certain treatment problems CD patients are facing \cite{Wang2009}.

Zhong \textit{et al.} \cite{Zhong2010} proposed a modification to the pathway-based approach presented by Wang \textit{et al.} \cite{Wang2007a}. They use  well-defined SNPs associated with gene expression levels in different type of tissues.  These SNPs result from studies about the genetics of gene expressions (GGE) instead of studies about disease associations \cite{Zhong2010}. The SNPs are tested for associations with Type 2 Diabetes (T2D) in a GWAS. The associated SNPs are then mapped to genes from pathways available in KEGG database. Finally ES scores are calculated and highly enriched gene set are linked to their corresponding pathways \cite{Zhong2010}.

Aiming to identify overlapping and over-represented pathways from different studies about T2D etiology, Elbers \textit{et al.}  proposed the use of a pathway-based approach \cite{Elbers2009}. They combine the information of SNPs associated with T2D discovered in different GWA studies into one dataset in order to find common associations missed by the original GWA studies. These SNPs are then mapped to LD blocks, which contain several genes. The authors defined the LD blocks as the groups of SNPs that have a correlation coefficient greater than a certain threshold \cite{Elbers2009}.  SNPs that are successfully mapped to LD blocks with genes are included in the pathway-based study whereas SNPs mapped to blocks with no genes are excluded \cite{Elbers2009}. The authors then use a gene network tool called \textit{Prioritizer} in order to build functional gene networks using a bayesian approach and based on prior knowledge obtained from publicly available databases. The built gene networks consist of genes functionally related to other genes that contain SNPs in association with T2D.  Consequently, these networks offer the opportunity to assess the susceptibility to T2D affected by genes that would have been missed without searching for associations using these gene networks \cite{Elbers2009}. Finally, they compared five pathway-classification tools (KEGG, GATHER, DAVID, PANTHER, BioCarta) to check the over-representation of some of the susceptibility genes in the identified pathways.

Similarly, to asses and identify the polygenic basis of seven common diseases (bipolar disorder, coronary artery disease, Crohn’s disease , hypertension, rheumatoid arthritis, type 1 diabetes, and type 2 diabetes), Torkamani \textit{et al.} adopted a pathway-based approach to further analyze the Wellcome Trust Case Control Consortium (WTCCC) GWAS results \cite{Torkamani2008}.  The SNPs associated with the seven diseases are assigned to the their nearest gene. The highest SNP \textit{p-value} on each gene is considered as the gene's weight.  Then using the \textit{MetaCore} software, the identified genes are tested for enrichment in Maps, Diseases, GO processes and GeneGO to determine the biological pathways and networks shared by the seven diseases \cite{Torkamani2008}.

Lebrec \textit{et al.} presented a Bayesian linear model to integrate GWAS results with pathway and gene annotation data from GO \cite{Lebrec2009}.  The purpose of their approach is to use external sources of biological information in prioritizing candidate genes instead of using only GWAS results for prioritization.  The authors use the same approach proposed by Wang \textit{et al.} \cite{Wang2007a} to map associated SNPs to genes. They then build a matrix $X$ of size $n \times p$  where $n$ is the number of genes, $p$ is the number of pathways and $X_{ij}$ is set to one if gene $i$ belongs to the pathway $p$, and to zero otherwise. Afterwards, they model the effect of gene $i$ noted as $\mu_i$ on the disease of interest using their proposed approximate model \cite{Lebrec2009}.

\begin{figure}[!htp]
 \centering
   \includegraphics[scale=.6]{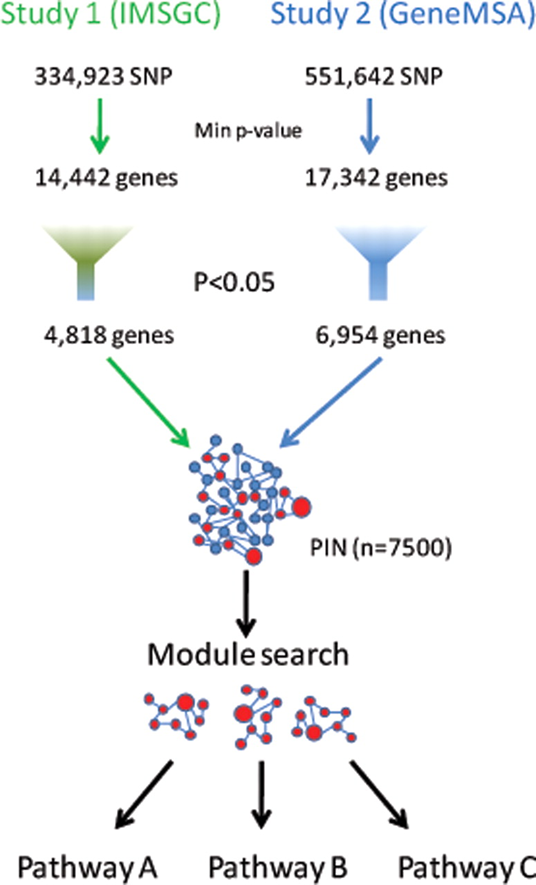}
  \caption{ A diagram obtained from Baranzini \textit{et al.} \cite{Baranzini2009}. It summarizes the proposed framework for the integration of network-based pathway analysis with GWAS.}
\label {fig:ms}
\end{figure}

Baranzini et. al introduced a network-based pathway analysis for Multiple Sclerosis (MS) \cite{Baranzini2009}. Their work is based on the hypothesis that genes containing SNPs with modest effect on disease can be identified in case they all belong to the same biological pathway or mechanism \cite{Baranzini2009}. Their analysis adopts a protein interaction network (PIN) analysis with GWAS. Two GWA studies are performed on two different SNP data published by Baranzini et al. and the International Multiple Sclerosis Genetics Consortium \cite{Baranzini2009a, Consortium2007}, where \textit{p-value}s are calculated for SNPs demonstrating association with MS. Then SNPs are filtered according to a \textit{p-value} threshold of less than 0.05, and the ones with \textit{p-values} less than the threshold are included in further analysis. Disease-associated SNPs are mapped to genes, and their \textit{p-value}s are used by the well-known open-source software called \textit{Cytoscape} plugin \textit{jActive modules} to extract the protein interaction network \cite{Baranzini2009}. Afterwards, sub-networks of interacting gene products that are associated with MS are identified. Finally, significant gene modules are included in an enrichment test using KEGG pathways. The proposed framework  is summarized in Figure~\ref{fig:ms}, obtained from the publication of Baranzini \textit{et al.} \cite{Baranzini2009}.
 
The methods presented above use a variety of pathway databases. Authors usually choose one database over another according to their familiarity with the database and sometimes opt to combine pathway information from different databases. However, pathway information differs from one database to another, for example Notch signaling pathway in KEGG is different from the one in Reactome as the entities involved in this pathway are not identical in both pathway databases. Therefore, the results of a study using pathway information from a certain database will not necessarily be validated if another database is used. Hence, there is a need to provide either a mapping scheme between different pathway databases or a standardized presentation of pathways in a centralized database in order to ensure consistency in GWAS using pathway analysis.

\section{Interaction and Function Information}
In this section, the surveyed approaches focus on the integration of genes/markers interaction and functional information with GWAS. The purpose of this integration is to increase the power of GWAS by decreasing the burden of multiple association testing when providing means to group variants in sets according to previously reported knowledge regarding their interactions and functionality.  Incorporating additional resources of information enables performing  a more comprehensive analysis in GWAS that leads to increased evidence in reported associations  beyond the sheer evidence of \textit{p-values}. The literature reviewed in this section varies in the type of the integrated information, which is either: gene-gene interactions (epistasis) or functional relatedness of genes or SNPs functionality \cite{Pan2008}.

Epistasis is an effect in which some genes mask other genes\rq{} contribution to certain phenotypes. This gene-gene interaction has an important role in disease risk and poses a great challenge to GWAS.  Analyzing the combinatorial effect of pairs of marker/genes on disease, to account for epistasis, requires a tremendous amount of computational power \cite{Bush2009}. However, the literature contains a significant amount of information regarding epistasis. Therefore, integrating previously acquired knowledge about gene interactions offers an opportunity to group SNPs related through gene epistasis and increase the evidence about associations with modest/rare effects.  Bush \textit{et al.} \cite{Bush2009} proposed a system  of biological knowledge integration,  called Biofilter, where multi-SNPs (a group SNPs) are modeled and weighted before being used in GWAS. It integrates various sources of data that offer information about gene interactions, including pathways, protein-protein interactions and gene expression. All the studied SNPs  are mapped to their corresponding genes. Then data about  interactions of gene pairs is collected from publicly available databases such as GO, KEGG and Reactome. Relationships between multiple SNPs are inferred from their genes interaction information using pair-wise combinations of SNP-related-genes and these relationships are called ``multi-SNP models". Each multi-SNP model is weighed according to the proposed \lq\lq{}implication index\rq\rq{}, which quantifies the knowledge base support \cite{Bush2009}. The implication index is calculated by summing up  the number of databases that provide evidence of gene$\times$gene interactions. Finally, weighted multi-SNP models are analyzed in a GWAS to uncover the associations between the disease of interest and these models.

Due to gene-gene interactions and the fact that complex diseases are influenced by multiple genes, the analysis of risk factors necessitates using multilocus analysis approaches \cite{Pan2008}. With the purpose of reducing the cost of multiple testing, Pan proposes a gene network-based weighing model for detecting disease association with multiple loci. Pan\rq{}s approach assumes that interactions between gene products (protein protein interaction (PPI)) is a sign of interactions between genes themselves and if one of the interacting genes is associated with a disease, the other should also be associated. In Pan's study, genes included in the analysis are ranked according to the probability of being a disease-causing gene as reported by recent studies. Using PPIs information,  networks of the most probable disease-causing genes are built. Following the aforementioned assumption and using a network of 23 interacting genes, a network of disease-causing genes (DCG) is formed. According to the assumption that disease-causing genes tend to be closely interconnected in gene-networks,  the network of DCG is extended by adding direct neighbors (first-order neighbors, 1ON) of these genes according to their PPIs.  The idea of first order neighbors is then generalized to K-order neighbors ($k$ON) where higher order neighbors are included to form larger groups of DCG. $k$ON will be used to perform a multi-marker based GWAS, which will reduce the number of association tests.  

Wu \textit{et al.}  proposed an approach based on a distance measure to group SNPs into SNP-sets \cite{Wu2010}. The proximity of SNPs  to genomic features, such as genes, provides a measure for forming the SNP-set, which is then used in GWA multilocus testing. Integrating gene proximity information with GWAS, as a criterion to form multi-SNP groups, helps in decreasing the effect of multiple testing and enables the detection of the influence of epistasis interactions on disease risk \cite{Wu2010}.

As a post-processing step, Sohns \textit{et al.} suggest ranking genes of associated genetic markers using  a combination of Gene Set Enrichement Analysis (GSEA) and Hierarchical Bayes Prioritization (HBP) \cite{Sohns2009}. The two combined approaches aim to increase the power of selecting genes associated with the disease of interest. GSEA is a method originally developed for gene expression analysis and its application on GWAS data has been explained in Section~\ref{pathway}. In Sohns \textit{et al.}'s work,  HBP uses prior covariates of each variant in the dataset to rank these variants. Two types of regression models are applied to estimate the relationship between the prior covariates and the observed association statistic of variants, then the estimated regression coefficients are used to rank the variants. The first is a logistic model for variant-prior probabilities , which are calculated from variant distance to genes and gene set information. The second is a linear model applied to estimate association strength \cite{Sohns2009}. 

Despite their ability to increase the power of GWAS, the reviewed approaches are computationally expensive and require a significant amount of data preprocessing. In addition, there is a lack of a discussion about the scalability of these approaches especially when attempting to apply them on datasets with larger genetic coverage.

\section{Gene Expression Analyses}
Microarrays experiments are used to measure gene-expression levels. The starting point in gene expression analysis is finding the genes that are differentially expressed \cite{Goeman2007}. Next, the analysis usually proceeds to find patterns and clusters among differentially expressed group of genes. Gene-expression analysis has been an active research area for the past two decades utilizing myriads of statistical and machine learning methods \cite{Goeman2007}. With the goal of identifying novel genes pertaining to disease risk, gene-expression analysis is combined with GWA studies.  In this section, we discuss the use of gene expression data as complementary information to SNP data or as a pre/post processing of GWA data . 

Naukkarinen \textit{et al.}  explained the importance of integrating genome-wide expression data with SNPs data in order to find the missing heritability in disease \cite{Naukkarinen2010}. The authors  proposed a framework of three stages to integrate gene expression analysis with an association study on obesity and Body Mass Index (BMI) \cite{Naukkarinen2010}. The first stage consists of performing a gene expression analysis to identify the genes differentially expressed in tissues related to BMI. In the second stage, genes correlated with obesity are identified using a stringent Pearson's correlation threshold. From the first two stages, two groups of genes are formed: genes that are both differentially expressed and correlated with obesity are referred to as \textit{reactive genes} and genes that are only correlated with obesity are considered \textit{putative causative} genes. Both groups of genes are analyzed in the last stage where a GWA study is performed on the SNPs located within these genes. Then the GWAS results are replicated using the data of the GenMets study (a previously published GWAS about the genetics of metabolic syndrome) to increase the confidence in the reported associations. The integration approach proposed by Naukkarinen \textit{et al.} forms here a preprocessing step of GWAS data that limits the analysis to SNPs within differentially expressed genes.

Gene-expression analysis can also be integrated with GWAS as a post-processing approach to increase the evidence in the reported associations or highlight novel ones. As a post-processing step, Hsu \textit{et al.} proposed  using gene-expression analysis to further analyze the results obtained by a GWA study for Osteoporosis \cite{Hsu2010}. They first performed a GWAS analysis using a regression model, then replicated the top associations in two independent cohorts. The replication phase generated two sets of results, the first is the group of significantly associated loci and the second is the group of suggestive association loci that were selected according to two \textit{p-value}s thresholds. Reported suggestive association loci were prioritized according to their differential expression in Osteoporosis related tissues. The top prioritized genes in this set of suggestive association loci and the set of significantly associated loci are used in a GSEA to find if the selected genes are over-represented in certain biological/functional pathway then the gene-expression abundance of these genes is measured \cite{Hsu2010}.

Murphy \textit{et al.} performed an study to uncover the relationship between genome-wide SNP data and gene expression levels using linear models  \cite{Murphy2010}. As a consequence, SNPs that have strong association with the abundance of expression of gene, which are referred to as expression SNPs (eSNPs),  are mapped to their corresponding genes, and a gene set is formed \cite{Zhong2010}. The gene-expression levels were measured for blood lymphocytes. Using the associated gene set, an enrichment test was performed to compare the associated genes with previously curated catalog of GWAS of common diseases \cite{Murphy2010}. Similarly, Zeller \textit{et al.} also undertook an association study between genome-wide SNP data and the Transcriptome of monocytes using ANOVA \cite{Zeller2010}. Afterwards, the associations between genes that are differentially expressed and cardiovascular disease were tested using linear regression models. Finally, SNPs associated with differentially expressed genes were mapped to their corresponding genes and their functional classification was performed using ontology analysis \cite{Zeller2010}.

\begin{figure}[tb!]
 \centering
   \includegraphics[scale=.5]{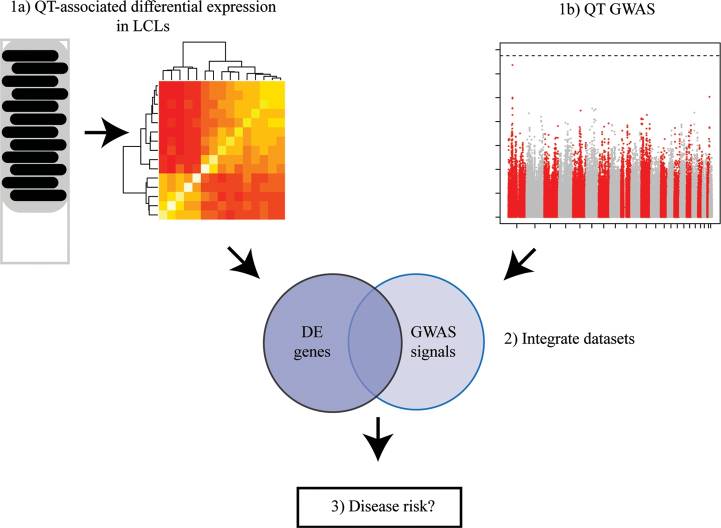}
  \caption{ A diagram obtained from a publication by Cusanovich \textit{et al.}[20]. It summarizes the flow of gene expression analysis (in 1a)) , GWAS (in 1b)) and the integration of both results (in 2)) to identify novel associations (in 3)) that would have been missed if GWAS alone was performed.  }
\label {fig:cusa}
\end{figure}

Aiming to identify novel genetic associations with small effects, Cusanovich \textit{et al.} proposed a complimentary approach to combine GWAS results with gene expression analysis in Lymphoblastoid cell lines (LCLs) for individuals suffering from Asthma \cite{Cusanovich2012}. The authors identified differentially expressed  (DE) genes for individuals that have high and low lymphocyte count (a physiological Quantitative Trait (QT) associated with asthma). In parallel, they explored the associations between genome-wide SNP data and lymphocyte count and mapped the associated SNPs to their corresponding genes. Consequently, two sets of genes resulting from the two previous analyses were integrated to identify the genes that are related to asthma and at the same time were missed by GWAS alone. Finally the intersection of these two sets of genes   was labeled as a disease risk after performing an enrichment test and comparing it to previous asthma GWA studies \cite{Cusanovich2012}. The flow of analyses proposed by Cusanovich \textit{et al.} \cite{Cusanovich2012} is summarized in Figure~\ref{fig:cusa}.

Gene expression data used in the work described above pertains to genes/tissues that have been previously reported to be related to the disease being studied. Nowadays myriads of sources of information about various complex diseases are available. Research oriented towards using gene expression levels measured within gene/tissue of diseases that share similar symptoms and/or pathways, provides an opportunity to reveal novel disease-associations that were difficult to find when examining gene expression data related only to one disease.

\chapter{Conclusion and Future Work}
In this report, we discussed the importance of association studies as a tool to reveal disease risk factors. The discussion was focused on genome-wide association studies (GWAS) since the state of the art in genotyping technologies enables the acquisition of genome-wide SNP data. We presented the history of genetic screening and association studies, which started with Fisher's model of disease susceptibility and has been transformed into GWAS.  In addition, we differentiated between the various types of association studies and their different design methods and data. The standard statistical approaches used in GWAS were also described and compared. Moreover, the challenges currently faced in GWAS were discussed along with the ways to address them.

The major goal of this report is to present the different approaches designed and implemented to improve GWA analysis and to increase its ability to discover disease risk factors. We discussed the research oriented towards modifying and improving the techniques used for the discovery of disease associations. Moreover, we explained several alternative approaches used for the significance evaluation of disease associations.  We also presented different recommendations about the design of GWA studies and discussed several approaches that aim to increase the genetic coverage and discard unwanted structure pertaining to population stratification or unreported individual relatedness in GWAS data.  Furthermore, we reviewed the literature that discusses integration of external information sources with GWAS. Incorporating additional information can be done either as preprocessing of GWA data or as post-processing. We discussed the four different integration approaches: meta-analyses, pathway analysis, gene expression analysis and interaction and function information.

Despite the wide range of proposed approaches to improve GWAS approaches, the discovery of disease risk factor of common complex disease is still far from complete. The acquired knowledge about current association loci explains a small portion of phenotypical and heritable disease variation. In addition, the translation of discovered association into clinical knowledge that answers questions about the etiology and chain of mechanisms of complex disease remains unclear. Therefore, the improvement of current GWAS approaches requires additional effort in taking advantage of the plethora of biological, physiological and medical information.  

 The continuing advancement in biotechnology point out the necessity to adopt scalable improvement approaches. Current GWA data spans hundreds of thousands to million of genetic markers. Yet, efforts to increase the genetic coverage will lead to  even larger GWA datasets, which also contain rarer genetic variants. Therefore, GWAS improvement approaches should be robust enough to detect rare disease associations and the search for associations  is supposed to be able to scale well with the increasing size of genetic datasets.  
 
 Research in the biomedical field provides a wide spectrum of information sources. The number of publicly available databases,  which contain sequences, gene expression, tissue information, protein-protein interactions and pathways, is increasing rapidly. Novel GWAS  approaches should be designed to make use of the variety of available information in order to increase the ability of GWAS to discover novel associations and to boost the confidence in the discovered associations. The integration of different information sources into the analysis of GWAS is a key factor towards bridging the gap between GWAS discoveries and their application in the clinical world.
   
The majority of surveyed  approaches were based on statistical analysis except for a few association testing approaches, which were based on machine learning and data mining techniques. Most of these statistical techniques propose the use of complex models to be able to reflect all different aspects of disease risk. However, the model complexity jeopardizes the scalability of the approach and increases the computational power needed for applying the model. New approaches should be designed to improve GWA studies while taking in consideration the scalability, coverage and integrability issues. Literature in computer science and machine learning contains extensive amount of research addressing pattern recognition, data mining and signal processing. Using this research provides a great opportunity  to improve GWAS approaches. Future research in GWAS is supposed to take advantage of the state of the art of computational methods that are significantly scalable, efficient and effective.

A future direction in GWAS is the consolidation of more than one of the approaches discussed in this report.   For instance, it is important to address the fact that the presented meta-analysis do not use pathway, functional and gene expression information to prioritize the previously reported GWAS results. Currently, we have access to a GWAS data of 500K SNPs for 1857 patients of Multiple Sclerosis (MS). In the catalog of published genome-wide association studies, there are numerous reported MS-SNP associations that can be harvested and prioritized in order to serve as prior knowledge for a new MS GWAS.  

Our future work plan is implement a two-stage MS GWAS inspired by recent work of Kostem and Eskin \cite{Kostem2013}. They proposed to accelerate the process of association discovery using a two stage testing framework. First, they select a group of SNPs, referred to as proxy SNPs, by searching each LD region for the group of proxies that minimizes the number of required tests for GWAS. In each LD region, they calculate the association statistics for all proxies. If these statistics are lower than a certain threshold they test the remainder SNPs in this region, otherwise the region is discarded.  Our proposed work-flow incorporates an integrative meta-analysis and a selective association testing approach. The integrative meta-analysis will collect the previously reported MS associations and prioritize them according to the relationship between their genomic locations and well known MS related pathways. These prioritized associations provide information about important genomic locations that are likely to hold more associations with MS. As a consequence, the 500k MS SNP data will be partitioned into blocks of potential areas of association according to the genomic locus of the previously reported prioritized associations. As for the selective association testing approach, the SNPs belonging to the blocks containing the highly prioritized previous associations will be included in the set of SNPs to be tested for association with MS. 

Our proposed two-stage study tackles the lack of integration of pathway analysis into meta-analysis data. In addition, it provides a testing approach that minimizes the computational and time complexity of GWAS by limiting the testing process to only potential areas of associations. Instead of exhaustively testing all SNPs in the 500K dataset, our proposed approach is to test a SNP only if this testing step is supported by previously reported association results and pathway information. Thus, our future work plan builds on previously published studies and takes advantage of publicly available knowledge about MS in order to increase the chance of discovering disease-causal variants/genes. As a consequence, our approach contributes to the solution of the missing heritability problem in MS and facilitates a clearer  understanding of this disease's etiology and mechanism of actions. 

\clearpage

\bibliographystyle{plain}%\nocite{*}	
\bibliography{depth_proposal_refs}

\begin{thebibliography}{100}

\bibitem{Armitage1955}
P.~Armitage.
\newblock Tests for linear trends in proportions and frequencies.
\newblock {\em Biometrics}, 11(3):375--386, 1955.

\bibitem{Bakker2005}
P.~I. W.~De Bakker, R.~Yelensky, I.~Pe'er, S.~B. Gabriel, M.~J. Daly, and
  D.~Altshuler.
\newblock Efficiency and power in genetic association studies.
\newblock {\em Nature genetics}, 37(11):1217--1223, 2005.

\bibitem{Baranzini2009}
S.~E. Baranzini, N.~W. Galwey, J.~Wang, P.~Khankhanian, R.~Lindberg,
  D.~Pelletier, W.~Wu, B.~M.~J. Uitdehaag, L.~Kappos, and C.~H. Polman.
\newblock Pathway and network-based analysis of genome-wide association studies
  in multiple sclerosis.
\newblock {\em Human molecular genetics}, 18(11):2078, 2009.

\bibitem{Baranzini2009a}
Sergio~E. Baranzini, Joanne Wang, Rachel~A. Gibson, Nicholas Galwey, Yvonne
  Naegelin, Frederik Barkhof, Ernst-Wilhelm Radue, Raija~LP Lindberg,
  Bernard~MG Uitdehaag, and Michael~R. Johnson.
\newblock Genome-wide association analysis of susceptibility and clinical
  phenotype in multiple sclerosis.
\newblock {\em Human molecular genetics}, 18(4):767--778, 2009.

\bibitem{Benjamini1995}
Y.~Benjamini and Y.~Hochberg.
\newblock Controlling the false discovery rate: a practical and powerful
  approach to multiple testing.
\newblock {\em Journal of the Royal Statistical Society.Series B
  (Methodological)}, pages 289--300, 1995.

\bibitem{Berger1987}
James~O. Berger and Thomas Sellke.
\newblock Testing a point null hypothesis: the irreconcilability of p values
  and evidence.
\newblock {\em Journal of the American Statistical Association},
  82(397):112--122, 1987.

\bibitem{Bertram2007}
L.~Bertram, M.~B. McQueen, K.~Mullin, D.~Blacker, and R.~E. Tanzi.
\newblock Systematic meta-analyses of alzheimer disease genetic association
  studies: the alzgene database.
\newblock {\em Nature genetics}, 39(1):17--24, 2007.

\bibitem{Bravo2010}
Hector~Corrada Bravo and Rafael~A. Irizarry.
\newblock Model-based quality assessment and base-calling for second-generation
  sequencing data.
\newblock {\em Biometrics}, 66(3):665--674, 2010.

\bibitem{Brookes2001}
Anthony~J Brookes.
\newblock {\em Single Nucleotide Polymorphism (SNP)}.
\newblock John Wiley \& Sons, Ltd, 2001.

\bibitem{Burton2007}
Paul~R. Burton, David~G. Clayton, Lon~R. Cardon, Nick Craddock, Panos Deloukas,
  Audrey Duncanson, Dominic~P. Kwiatkowski, Mark~I. McCarthy, Willem~H.
  Ouwehand, and Nilesh~J. Samani.
\newblock Genome-wide association study of 14,000 cases of seven common
  diseases and 3,000 shared controls.
\newblock {\em Nature}, 447(7145):661--678, 2007.

\bibitem{Bush2009}
W.~S. Bush, S.~M. Dudek, and M.~D. Ritchie.
\newblock Biofilter: a knowledge-integration system for the multi-locus
  analysis of genome-wide association studies.
\newblock In {\em Pacific Symposium on Biocomputing. Pacific Symposium on
  Biocomputing}, page 368. NIH Public Access, 2009.

\bibitem{Cantor2010}
R.~M. Cantor, K.~Lange, and J.~S. Sinsheimer.
\newblock Prioritizing gwas results: A review of statistical methods and
  recommendations for their application.
\newblock {\em The American Journal of Human Genetics}, 86(1):6--22, 2010.

\bibitem{Chuang2009}
L.~Y. Chuang, Y.~J.~Hou Jr, and C.~H. Yang.
\newblock A novel prediction method for tag snp selection using genetic
  algorithm based on knn.
\newblock {\em World Academy of Science, Engineering and Technology}, 53, 2009.

\bibitem{Clarke2011}
G.~M. Clarke, C.~A. Anderson, F.~H. Pettersson, L.~R. Cardon, A.~P. Morris, and
  K.~T. Zondervan.
\newblock Basic statistical analysis in genetic case-control studies.
\newblock {\em Nature protocols}, 6(2):121--133, 2011.

\bibitem{Comings2001}
David~E Comings.
\newblock Polygenic disorders.
\newblock In {\em eLS}. John Wiley \& Sons, Ltd, 2001.

\bibitem{Consortium2007}
International Multiple Sclerosis~Genetics Consortium, D.~A. Hafler,
  A.~Compston, S.~Sawcer, E.~S. Lander, M.~J. Daly, P.~L.~De Jager, P.~I.
  de~Bakker, S.~B. Gabriel, D.~B. Mirel, A.~J. Ivinson, M.~A. Pericak-Vance,
  S.~G. Gregory, J.~D. Rioux, J.~L. McCauley, J.~L. Haines, L.~F. Barcellos,
  B.~Cree, J.~R. Oksenberg, and S.~L. Hauser.
\newblock Risk alleles for multiple sclerosis identified by a genomewide study.
\newblock {\em The New England journal of medicine}, 357(9):851--862, Aug 30
  2007.

\bibitem{Craddock2010}
N.~Craddock, M.~E. Hurles, N.~Cardin, R.~D. Pearson, V.~Plagnol, S.~Robson,
  D.~Vukcevic, C.~Barnes, D.~F. Conrad, and E.~Giannoulatou.
\newblock Genome-wide association study of cnvs in 16,000 cases of eight common
  diseases and 3,000 shared controls.
\newblock {\em Nature}, 464(7289):713--720, 2010.

\bibitem{Cusanovich2012}
D.~A. Cusanovich, C.~Billstrand, X.~Zhou, C.~Chavarria, S.~De Leon,
  K.~Michelini, A.~A. Pai, C.~Ober, and Y.~Gilad.
\newblock The combination of a genome-wide association study of lymphocyte
  count and analysis of gene expression data reveals novel asthma candidate
  genes.
\newblock {\em Human molecular genetics}, 2012.

\bibitem{Davidovich2009}
O.~Davidovich, G.~Kimmel, E.~Halperin, and R.~Shamir.
\newblock Increasing the power of association studies by imputation-based
  sparse tag snp selection.
\newblock {\em Communications in Information \& Systems}, 9(3):269--282, 2009.

\bibitem{Elbers2009}
C.~C. Elbers, K.~R. van Eijk, L.~Franke, F.~Mulder, Y.~T. van~der Schouw,
  C.~Wijmenga, and N.~C. Onland-Moret.
\newblock Using genome-wide pathway analysis to unravel the etiology of complex
  diseases.
\newblock {\em Genetic epidemiology}, 33(5):419--431, 2009.

\bibitem{Emslie2001}
Carol Emslie and Kate Hunt.
\newblock {\em Genetic Susceptibility}.
\newblock John Wiley \& Sons, Ltd, 2001.

\bibitem{Fallin2001}
D.~Fallin, A.~Cohen, L.~Essioux, I.~Chumakov, M.~Blumenfeld, D.~Cohen, and
  N.~J. Schork.
\newblock Genetic analysis of case/control data using estimated haplotype
  frequencies: application to apoe locus variation and alzheimer's disease.
\newblock {\em Genome research}, 11(1):143--151, 2001.

\bibitem{Feuk2006}
L.~Feuk, A.~R. Carson, and S.~W. Scherer.
\newblock Structural variation in the human genome.
\newblock {\em Nature Reviews Genetics}, 7(2):85--97, 2006.

\bibitem{Fisher1918}
Ronald Fisher.
\newblock The correlation between relatives on the supposition of mendelian
  inheritance.
\newblock {\em Philosophical Transactions of the Royal Society of
  Londonranscations of the Royal Society of Edinburgh}, 52:399--433, 1918.

\bibitem{Gibbs2003}
R.~A. Gibbs, J.~W. Belmont, P.~Hardenbol, T.~D. Willis, F.~Yu, H.~Yang, L.~Y.
  Ch'ang, W.~Huang, B.~Liu, and Y.~Shen.
\newblock The international hapmap project.
\newblock {\em Nature}, 426(6968):789--796, 2003.

\bibitem{Gibson2010}
G.~Gibson.
\newblock Hints of hidden heritability in gwas.
\newblock {\em Nat Genet}, 42(7):558--560, 2010.

\bibitem{Goeman2007}
J.~J. Goeman and P.~Buhlmann.
\newblock Analyzing gene expression data in terms of gene sets: methodological
  issues.
\newblock {\em Bioinformatics}, 23(8):980--987, 2007.

\bibitem{Goh2007}
Kwang-Il Goh, Michael~E. Cusick, David Valle, Barton Childs, Marc Vidal, and
  Albert-Laszlo Barabasi.
\newblock The human disease network.
\newblock {\em Proceedings of the National Academy of Sciences},
  104(21):8685--8690, 2007.

\bibitem{Hemminki2006}
K.~Hemminki, J.~L. Bermejo, and A.~Forsti.
\newblock The balance between heritable and environmental aetiology of human
  disease.
\newblock {\em Nature Reviews Genetics}, 7(12):958--965, 2006.

\bibitem{Hemminki2008}
K.~Hemminki, A.~Forsti, and J.~L. Bermejo.
\newblock The common disease-common variant hypothesis and familial risks.
\newblock {\em PLoS one}, 3(6):e2504, 2008.

\bibitem{Hsu2010}
Y.~H. Hsu, M.~C. Zillikens, S.~G. Wilson, C.~R. Farber, S.~Demissie,
  N.~Soranzo, E.~N. Bianchi, E.~Grundberg, L.~Liang, and J.~B. Richards.
\newblock An integration of genome-wide association study and gene expression
  profiling to prioritize the discovery of novel susceptibility loci for
  osteoporosis-related traits.
\newblock {\em PLoS genetics}, 6(6):e1000977, 2010.

\bibitem{Hunter2007}
D.~J. Hunter, P.~Kraft, K.~B. Jacobs, D.~G. Cox, M.~Yeager, S.~E. Hankinson,
  S.~Wacholder, Z.~Wang, R.~Welch, and A.~Hutchinson.
\newblock A genome-wide association study identifies alleles in fgfr2
  associated with risk of sporadic postmenopausal breast cancer.
\newblock {\em Nature genetics}, 39(7):870--874, 2007.

\bibitem{Ioannidis2009}
J.~P.~A. Ioannidis, G.~Thomas, and M.~J. Daly.
\newblock Validating, augmenting and refining genome-wide association signals.
\newblock {\em Nature Reviews Genetics}, 10(5):318--329, 2009.

\bibitem{Ji2008}
W.~Ji, J.~N. Foo, B.~J. O'Roak, H.~Zhao, M.~G. Larson, D.~B. Simon, and
  C.~Newton-Cheh.
\newblock Rare independent mutations in renal salt handling genes contribute to
  blood pressure variation.
\newblock {\em Nature genetics}, 40(5):592--599, 2008.

\bibitem{Kang2010}
H.~M. Kang, J.~H. Sul, N.~A. Zaitlen, S.~Kong, N.~B. Freimer, C.~Sabatti, and
  E.~Eskin.
\newblock Variance component model to account for sample structure in
  genome-wide association studies.
\newblock {\em Nature genetics}, 42(4):348--354, 2010.

\bibitem{Kang2008}
Hyun~Min Kang, Noah~A. Zaitlen, Claire~M. Wade, Andrew Kirby, David Heckerman,
  Mark~J. Daly, and Eleazar Eskin.
\newblock Efficient control of population structure in model organism
  association mapping.
\newblock {\em Genetics}, 178(3):1709--1723, 2008.

\bibitem{Keating2008}
Brendan~J. Keating, Sam Tischfield, Sarah~S. Murray, Tushar Bhangale, Thomas~S.
  Price, Joseph~T. Glessner, Luana Galver, Jeffrey~C. Barrett, Struan~FA Grant,
  and Deborah~N. Farlow.
\newblock Concept, design and implementation of a cardiovascular gene-centric
  50 k snp array for large-scale genomic association studies.
\newblock {\em PloS one}, 3(10):e3583, 2008.

\bibitem{Khoury2009}
M.~J. Khoury, L.~Bertram, P.~Boffetta, A.~S. Butterworth, S.~J. Chanock, S.~M.
  Dolan, I.~Fortier, M.~Garcia-Closas, M.~Gwinn, and J.~Higgins.
\newblock Genome-wide association studies, field synopses, and the development
  of the knowledge base on genetic variation and human diseases.
\newblock {\em American Journal of Epidemiology}, 170(3):269, 2009.

\bibitem{Klein2005}
R.~J. Klein, C.~Zeiss, E.~Y. Chew, J.~Y. Tsai, R.~S. Sackler, C.~Haynes, A.~K.
  Henning, J.~P. SanGiovanni, S.~M. Mane, and S.~T. Mayne.
\newblock Complement factor h polymorphism in age-related macular degeneration.
\newblock {\em Science}, 308(5720):385--389, 2005.

\bibitem{Kostem2013}
Emrah Kostem and Eleazar Eskin.
\newblock Efficiently identifying significant associations in genome-wide
  association studies.
\newblock In {\em Research in Computational Molecular Biology}, pages 118--131.
  Springer, 2013.

\bibitem{Kotsiantis2007}
SB~Kotsiantis, ID~Zaharakis, and PE~Pintelas.
\newblock Supervised machine learning: A review of classification techniques.
\newblock {\em Frontiers in Artificial Intelligence and Applications}, 160:3,
  2007.

\bibitem{Ku2010}
CS~Ku, EY~Loy, Y.~Pawitan, and KS~Chia.
\newblock The pursuit of genome-wide association studies: where are we now?
\newblock {\em Journal of human genetics}, 55(4):195, 2010.

\bibitem{Kwee2008}
L.~C. Kwee, D.~Liu, X.~Lin, D.~Ghosh, and M.~P. Epstein.
\newblock A powerful and flexible multilocus association test for quantitative
  traits.
\newblock {\em The American Journal of Human Genetics}, 82(2):386--397, 2008.

\bibitem{Lawrence2005}
E.~Lawrence.
\newblock {\em Henderson's dictionary of biology/[edited by] Eleanor Lawrence}.
\newblock Prentice Hall, 2005.

\bibitem{le2011}
B.~Nguyen le, S.~J. Diskin, M.~Capasso, K.~Wang, M.~A. Diamond, J.~Glessner,
  C.~Kim, E.~F. Attiyeh, Y.~P. Mosse, K.~Cole, A.~Iolascon, M.~Devoto,
  H.~Hakonarson, H.~K. Li, and J.~M. Maris.
\newblock Phenotype restricted genome-wide association study using a
  gene-centric approach identifies three low-risk neuroblastoma susceptibility
  loci.
\newblock {\em PLoS genetics}, 7(3):e1002026, Mar 2011.

\bibitem{Lebrec2009}
J.~Lebrec, T.~Huizinga, R.~Toes, J.~Houwing-Duistermaat, and H.~van
  Houwelingen.
\newblock Integration of gene ontology pathways with north american rheumatoid
  arthritis consortium genome-wide association data via linear modeling.
\newblock In {\em BMC proceedings}, volume~3, page S94. BioMed Central Ltd,
  2009.

\bibitem{Lee2011}
S.~H. Lee, N.~R. Wray, M.~E. Goddard, and P.~M. Visscher.
\newblock Estimating missing heritability for disease from genome-wide
  association studies.
\newblock {\em The American Journal of Human Genetics}, 2011.

\bibitem{Lewis2012}
C.~M. Lewis and J.~Knight.
\newblock Introduction to genetic association studies.
\newblock {\em Cold Spring Harbor Protocols}, 2012(3):pdb. top068163, 2012.

\bibitem{Li2009a}
M.~Li, K.~Wang, S.~F.~A. Grant, H.~Hakonarson, and C.~Li.
\newblock Atom: a powerful gene-based association test by combining optimally
  weighted markers.
\newblock {\em Bioinformatics}, 25(4):497--503, 2009.

\bibitem{Li2009}
Y.~Li, C.~Willer, S.~Sanna, and G.~Abecasis.
\newblock Genotype imputation.
\newblock {\em Annual review of genomics and human genetics}, 10:387, 2009.

\bibitem{Liu2010}
J.~Z. Liu, A.~F. Mcrae, D.~R. Nyholt, S.~E. Medland, N.~R. Wray, K.~M. Brown,
  N.~K. Hayward, G.~W. Montgomery, P.~M. Visscher, and N.~G. Martin.
\newblock A versatile gene-based test for genome-wide association studies.
\newblock {\em The American Journal of Human Genetics}, 87(1):139--145, 2010.

\bibitem{Lohmueller2003}
K.~E. Lohmueller, C.~L. Pearce, M.~Pike, E.~S. Lander, and J.~N. Hirschhorn.
\newblock Meta-analysis of genetic association studies supports a contribution
  of common variants to susceptibility to common disease.
\newblock {\em Nature genetics}, 33(2):177--182, 2003.

\bibitem{Manolio2009a}
T.~A. Manolio.
\newblock Collaborative genome-wide association studies of diverse diseases:
  programs of the nhgri's office of population genomics.
\newblock {\em Pharmacogenomics}, 10(2):235--241, 2009.

\bibitem{Manolio2009}
T.~A. Manolio, F.~S. Collins, N.~J. Cox, D.~B. Goldstein, L.~A. Hindorff, D.~J.
  Hunter, M.~I. McCarthy, E.~M. Ramos, L.~R. Cardon, and A.~Chakravarti.
\newblock Finding the missing heritability of complex diseases.
\newblock {\em Nature}, 461(7265):747--753, 2009.

\bibitem{Marchini2007}
J.~Marchini, B.~Howie, S.~Myers, G.~McVean, and P.~Donnelly.
\newblock A new multipoint method for genome-wide association studies by
  imputation of genotypes.
\newblock {\em Nature genetics}, 39(7):906--913, 2007.

\bibitem{McCarthy2008b}
M.~I. McCarthy, G.~R. Abecasis, L.~R. Cardon, D.~B. Goldstein, J.~Little,
  JP~Ioannidis, and J.~N. Hirschhorn.
\newblock Genome-wide association studies for complex traits: consensus,
  uncertainty and challenges.
\newblock {\em Nature Reviews Genetics}, 9(5):356--369, 2008.

\bibitem{McCarthy2008}
M.~I. McCarthy and J.~N. Hirschhorn.
\newblock Genome-wide association studies: potential next steps on a genetic
  journey.
\newblock {\em Human molecular genetics}, 17(R2):R156, 2008.

\bibitem{McElroy2008}
JP~McElroy and JR~Oksenberg.
\newblock Multiple sclerosis genetics.
\newblock {\em Advances in multiple Sclerosis and Experimental Demyelinating
  Diseases}, pages 45--72, 2008.

\bibitem{Miclaus2009}
Kelci Miclaus, Russ Wolfinger, and Wendy Czika.
\newblock Snp selection and multidimensional scaling to quantify population
  structure.
\newblock {\em Genetic epidemiology}, 33(6):488--496, 2009.

\bibitem{Moore2010}
J.~H. Moore, F.~W. Asselbergs, and S.~M. Williams.
\newblock Bioinformatics challenges for genome-wide association studies.
\newblock {\em Bioinformatics}, 26(4):445--455, 2010.

\bibitem{Morris1997}
AP~Morris, JC~Whittaker, and RN~Curnow.
\newblock A likelihood ratio test for detecting patterns of disease-marker
  association.
\newblock {\em Annals of Human Genetics}, 61(4):335--350, 1997.

\bibitem{Motulsky2006}
A.~G. Motulsky.
\newblock Genetics of complex diseases.
\newblock {\em Journal of Zhejiang University-Science B}, 7(2):167--168, 2006.

\bibitem{Murphy2010}
A.~Murphy, J.~H. Chu, M.~Xu, V.~J. Carey, R.~Lazarus, A.~Liu, S.~J. Szefler,
  R.~Strunk, K.~DeMuth, and M.~Castro.
\newblock Mapping of numerous disease-associated expression polymorphisms in
  primary peripheral blood cd4 lymphocytes.
\newblock {\em Human molecular genetics}, 19(23):4745, 2010.

\bibitem{Naukkarinen2010}
J.~Naukkarinen, I.~Surakka, K.~H. Pietilainen, A.~Rissanen, V.~Salomaa,
  S.~Ripatti, H.~Yki-Jarvinen, C.~M.~Van Duijn, H.~E. Wichmann, and J.~Kaprio.
\newblock Use of genome-wide expression data to mine the "gray zone" of gwa
  studies leads to novel candidate obesity genes.
\newblock {\em PLoS genetics}, 6(6):e1000976, 2010.

\bibitem{Nguyen2009}
Tuan~V. Nguyen.
\newblock Interpretation of randomized controlled trials of fracture
  prevention.
\newblock {\em IBMS BoneKEy}, 6(8):279--294, 2009.

\bibitem{Ohashi2003}
J.~Ohashi, S.~Yamamoto, N.~Tsuchiya, Y.~Hatta, T.~Komata, M.~Matsushita, and
  K.~Tokunaga.
\newblock Comparison of statistical power between 2x2 allele frequency and
  allele positivity tables in case/control studies of complex disease genes.
\newblock {\em Annals of Human Genetics}, 65(2):197--206, 2003.

\bibitem{Pan2008}
W.~Pan.
\newblock Network-based model weighting to detect multiple loci influencing
  complex diseases.
\newblock {\em Human genetics}, 124(3):225--234, 2008.

\bibitem{Park2011}
J.~H. Park, M.~H. Gail, C.~R. Weinberg, R.~J. Carroll, C.~C. Chung, Z.~Wang,
  S.~J. Chanock, J.~F.~Fraumeni Jr, and N.~Chatterjee.
\newblock Distribution of allele frequencies and effect sizes and their
  interrelationships for common genetic susceptibility variants.
\newblock {\em Proceedings of the National Academy of Sciences},
  108(44):18026--18031, 2011.

\bibitem{Pearson2008}
T.~A. Pearson and T.~A. Manolio.
\newblock How to interpret a genome-wide association study.
\newblock {\em JAMA: the journal of the American Medical Association},
  299(11):1335, 2008.

\bibitem{Peer2008}
I.~Pe'er, R.~Yelensky, D.~Altshuler, and M.~J. Daly.
\newblock Estimation of the multiple testing burden for genomewide association
  studies of nearly all common variants.
\newblock {\em Genetic epidemiology}, 32(4):381--385, 2008.

\bibitem{Ponder2001}
B.~A.~J. Ponder, G.~I. Evan, K.~H. Vousden, J.~Taipale, P.~A. Beachy,
  P.~Blume-jensen, T.~Hunter, J.~Hoeijmakers, L.~A. Liotta, and E.~C. Kohn.
\newblock Cancer genetics.
\newblock {\em Cancer}, 411(6835), 2001.

\bibitem{Price2006}
A.~L. Price, N.~J. Patterson, R.~M. Plenge, M.~E. Weinblatt, N.~A. Shadick, and
  D.~Reich.
\newblock Principal components analysis corrects for stratification in
  genome-wide association studies.
\newblock {\em Nature genetics}, 38(8):904--909, 2006.

\bibitem{Rakyan2011}
V.~K. Rakyan, T.~A. Down, D.~J. Balding, and S.~Beck.
\newblock Epigenome-wide association studies for common human diseases.
\newblock {\em Nature Reviews Genetics}, 12(8):529--541, 2011.

\bibitem{Risch1996}
N.~Risch and K.~Merikangas.
\newblock The future of genetic studies of complex human diseases.
\newblock {\em Science-AAAS-Weekly Paper Edition}, 273(5281):1516--1517, 1996.

\bibitem{Rokach2007}
Lior Rokach.
\newblock {\em Data mining with decision trees: theory and applications},
  volume~69.
\newblock World scientific, 2007.

\bibitem{Rosenberg2010}
N.~A. Rosenberg, L.~Huang, E.~M. Jewett, Z.~A. Szpiech, I.~Jankovic, and
  M.~Boehnke.
\newblock Genome-wide association studies in diverse populations.
\newblock {\em Nature Reviews Genetics}, 11(5):356--366, 2010.

\bibitem{Schadt2009}
E.~E. Schadt.
\newblock Molecular networks as sensors and drivers of common human diseases.
\newblock {\em Nature}, 461(7261):218--223, 2009.

\bibitem{Schork1997}
N.~J. Schork.
\newblock Genetics of complex disease.
\newblock {\em American journal of respiratory and critical care medicine},
  156(4):S103--S109, 1997.

\bibitem{Sieberts2007}
Solveig~K. Sieberts and Eric~E. Schadt.
\newblock Moving toward a system genetics view of disease.
\newblock {\em Mammalian Genome}, 18(6-7):389--401, 2007.

\bibitem{So2011}
H.~C. So, A.~H.~S. Gui, S.~S. Cherny, and P.~C. Sham.
\newblock Evaluating the heritability explained by known susceptibility
  variants: a survey of ten complex diseases.
\newblock {\em Genetic epidemiology}, 2011.

\bibitem{Sohns2009}
M.~Sohns, A.~Rosenberger, and H.~Bickeboller.
\newblock Integration of a priori gene set information into genome-wide
  association studies.
\newblock In {\em BMC proceedings}, volume~3, page S95. BioMed Central Ltd,
  2009.

\bibitem{Stephens2009}
M.~Stephens and D.~J. Balding.
\newblock Bayesian statistical methods for genetic association studies.
\newblock {\em Nature Reviews Genetics}, 10(10):681--690, 2009.

\bibitem{Storey2003}
J.~D. Storey and R.~Tibshirani.
\newblock Statistical significance for genomewide studies.
\newblock {\em Proceedings of the National Academy of Sciences of the United
  States of America}, 100(16):9440, 2003.

\bibitem{Subramanian2005}
Aravind Subramanian, Pablo Tamayo, Vamsi~K. Mootha, Sayan Mukherjee,
  Benjamin~L. Ebert, Michael~A. Gillette, Amanda Paulovich, Scott~L. Pomeroy,
  Todd~R. Golub, and Eric~S. Lander.
\newblock Gene set enrichment analysis: a knowledge-based approach for
  interpreting genome-wide expression profiles.
\newblock {\em Proceedings of the National Academy of Sciences of the United
  States of America}, 102(43):15545--15550, 2005.

\bibitem{Tenesa2008}
A.~Tenesa, S.~M. Farrington, J.~G.~D. Prendergast, M.~E. Porteous, M.~Walker,
  N.~Haq, R.~A. Barnetson, E.~Theodoratou, R.~Cetnarskyj, and N.~Cartwright.
\newblock Genome-wide association scan identifies a colorectal cancer
  susceptibility locus on 11q23 and replicates risk loci at 8q24 and 18q21.
\newblock {\em Nature genetics}, 40(5):631--637, 2008.

\bibitem{Thomas2009}
G.~Thomas, K.~B. Jacobs, P.~Kraft, M.~Yeager, S.~Wacholder, D.~G. Cox, S.~E.
  Hankinson, A.~Hutchinson, Z.~Wang, and K.~Yu.
\newblock A multistage genome-wide association study in breast cancer
  identifies two new risk alleles at 1p11. 2 and 14q24. 1 (rad51l1).
\newblock {\em Nature genetics}, 41(5):579--584, 2009.

\bibitem{Torkamani2008}
A.~Torkamani, E.~J. Topol, and N.~J. Schork.
\newblock Pathway analysis of seven common diseases assessed by genome-wide
  association.
\newblock {\em Genomics}, 92(5):265--272, 2008.

\bibitem{Turner2000}
N.~Turner.
\newblock Chi-squared test.
\newblock {\em J Clin Nurs}, 9:93, 2000.

\bibitem{Urbach2012}
D.~Urbach, M.~Lupien, M.~R. Karagas, and J.~H. Moore.
\newblock Cancer heterogeneity: origins and implications for genetic
  association studies.
\newblock {\em Trends in Genetics}, 2012.

\bibitem{Venter2001}
J.~C. Venter, M.~D. Adams, E.~W. Myers, P.~W. Li, R.~J. Mural, G.~G. Sutton,
  H.~O. Smith, M.~Yandell, C.~A. Evans, and R.~A. Holt.
\newblock The sequence of the human genome.
\newblock {\em Science Signalling}, 291(5507):1304, 2001.

\bibitem{Via2010}
M.~Via, C.~Gignoux, and E.~G. Burchard.
\newblock The 1000 genomes project: new opportunities for research and social
  challenges.
\newblock {\em Genome Med}, 2(3), 2010.

\bibitem{Wallace2009}
H.~M. Wallace.
\newblock {\em Genetic screening for susceptibility to disease}.
\newblock Wiley Online Library, 2009.

\bibitem{Wang2007a}
K.~Wang, M.~Li, and M.~Bucan.
\newblock Pathway-based approaches for analysis of genomewide association
  studies.
\newblock {\em The American Journal of Human Genetics}, 81(6):1278--1283, 2007.

\bibitem{Wang2010}
K.~Wang, M.~Li, and H.~Hakonarson.
\newblock Analysing biological pathways in genome-wide association studies.
\newblock {\em Nature Reviews Genetics}, 11(12):843--854, 2010.

\bibitem{Wang2009}
K.~Wang, H.~Zhang, S.~Kugathasan, V.~Annese, J.~P. Bradfield, R.~K. Russell,
  P.~Sleiman, M.~Imielinski, J.~Glessner, and C.~Hou.
\newblock Diverse genome-wide association studies associate the il12/il23
  pathway with crohn disease.
\newblock {\em The American Journal of Human Genetics}, 84(3):399--405, 2009.

\bibitem{Wang2011}
Lili Wang, Pouya Khankhanian, Sergio Baranzini, and Parvin Mousavi.
\newblock ictnet: A cytoscape plugin to produce and analyze integrative complex
  traits networks.
\newblock {\em BMC bioinformatics}, 12(1):380, 2011.

\bibitem{Wang2007}
T.~Wang and R.~C. Elston.
\newblock Improved power by use of a weighted score test for linkage
  disequilibrium mapping.
\newblock {\em The American Journal of Human Genetics}, 80(2):353--360, 2007.

\bibitem{Wellek2012}
S.~Wellek and A.~Ziegler.
\newblock Cochran-armitage test versus logistic regression in the analysis of
  genetic association studies.
\newblock {\em Human heredity}, 73(1):14--17, 2012.

\bibitem{Wilkie2001}
Andrew~OM Wilkie.
\newblock {\em Polygenic Inheritance and Genetic Susceptibility Screening}.
\newblock John Wiley \& Sons, Ltd, 2001.

\bibitem{Willer2010}
C.~J. Willer, Y.~Li, and G.~R. Abecasis.
\newblock Metal: fast and efficient meta-analysis of genomewide association
  scans.
\newblock {\em Bioinformatics}, 26(17):2190--2191, 2010.

\bibitem{Wu2010}
M.~C. Wu, P.~Kraft, M.~P. Epstein, D.~M. Taylor, S.~J. Chanock, D.~J. Hunter,
  and X.~Lin.
\newblock Powerful snp-set analysis for case-control genome-wide association
  studies.
\newblock {\em The American Journal of Human Genetics}, 86(6):929--942, 2010.

\bibitem{Wu2009}
T.~T. Wu, Y.~F. Chen, T.~Hastie, E.~Sobel, and K.~Lange.
\newblock Genome-wide association analysis by lasso penalized logistic
  regression.
\newblock {\em Bioinformatics}, 25(6):714--721, 2009.

\bibitem{Yasuno2010}
K.~Yasuno, K.~Bilguvar, P.~Bijlenga, S.~K. Low, B.~Krischek, G.~Auburger,
  M.~Simon, D.~Krex, Z.~Arlier, and N.~Nayak.
\newblock Genome-wide association study of intracranial aneurysm identifies
  three new risk loci.
\newblock {\em Nature genetics}, 42(5):420--425, 2010.

\bibitem{Zaitlen2005}
N.~A. Zaitlen, H.~M. Kang, M.~L. Feolo, S.~T. Sherry, E.~Halperin, and
  E.~Eskin.
\newblock Inference and analysis of haplotypes from combined genotyping studies
  deposited in dbsnp.
\newblock {\em Genome research}, 15(11):1594--1600, 2005.

\bibitem{Zeggini2009}
Eleftheria Zeggini and John~PA Ioannidis.
\newblock Meta-analysis in genome-wide association studies.
\newblock {\em Pharmacogenomics}, 10(2):191--201, 2009.

\bibitem{Zeller2010}
T.~Zeller, P.~Wild, S.~Szymczak, M.~Rotival, A.~Schillert, R.~Castagne,
  S.~Maouche, M.~Germain, K.~Lackner, and H.~Rossmann.
\newblock Genetics and beyond: the transcriptome of human monocytes and disease
  susceptibility.
\newblock {\em PLoS One}, 5(5):e10693, 2010.

\bibitem{Zhong2010}
H.~Zhong, X.~Yang, L.~M. Kaplan, C.~Molony, and E.~E. Schadt.
\newblock Integrating pathway analysis and genetics of gene expression for
  genome-wide association studies.
\newblock {\em The American Journal of Human Genetics}, 86(4):581--591, 2010.

\bibitem{Zhou2012}
X.~Zhou and M.~Stephens.
\newblock Genome-wide efficient mixed-model analysis for association studies.
\newblock {\em Nature genetics}, 44(7):821--824, 2012.

\end{thebibliography}
\end {document}